\documentstyle[prl,aps,epsf,twocolumn]{revtex}
\begin{document}
\draft

\def\i{\imath\,}
\def\ih{\frac{\imath}{2}\,}
\def\undertext#1{\vtop{\hbox{#1}\kern 1pt \hrule}}
\def\ra{\rightarrow}
\def\lfa{\leftarrow}
\def\ua{\uparrow}
\def\da{\downarrow}
\def\Ra{\Rightarrow}
\def\lra{\longrightarrow}
\def\ler{\leftrightarrow}
\def\lrb#1{\left(#1\right)}
\def\O#1{O\left(#1\right)}
\def\VEV#1{\left\langle\,#1\,\right\rangle}
\def\tr{\hbox{tr}\,}
\def\trb#1{\tr\lrb{#1}}
\def\dd#1{\frac{d}{d#1}}
\def\dbyd#1#2{\frac{d#1}{d#2}}
\def\pp#1{\frac{\partial}{\partial#1}}
\def\pbyp#1#2{\frac{\partial#1}{\partial#2}}
\def\ff#1{\frac{\delta}{\delta#1}}
\def\fbyf#1#2{\frac{\delta#1}{\delta#2}}
\def\pd#1{\partial_{#1}}
\def\br{\\ \nonumber & &}
\def\brr{\right. \\ \nonumber & &\left.}
\def\inv#1{\frac{1}{#1}}
\def\be{\begin{equation}}
\def\ee{\end{equation}}
\def\bea{\begin{eqnarray}}
\def\eea{\end{eqnarray}}
\def\ct#1{\cite{#1}}
\def\rf#1{(\ref{#1})}
\def\EXP#1{\exp\left(#1\right)}
\def\TEXP#1{\hat{T}\exp\left(#1\right)}
\def\INT#1#2{\int_{#1}^{#2}}
\def\MAT{{\it Mathematica }}
\def\LHS{left-hand side }
\def\RHS{right-hand side }
\def\COM#1#2{\left\lbrack #1\,,\,#2\right\rbrack}
\def\AC#1#2{\left\lbrace #1\,,\,#2\right\rbrace}

\twocolumn[\hsize\textwidth\columnwidth\hsize\csname@twocolumnfalse%
\endcsname

\preprint{}

\title{Fractionalization patterns in strongly correlated
electron systems: Spin-charge separation and beyond}
\author{Eugene Demler$^1$, Chetan Nayak$^2$,
Hae-Young Kee$^2$, Yong Baek Kim$^3$, and T. Senthil$^{4,5}$}
\address{$^1$Physics Department, Harvard University,
Cambridge, MA 02138\\
$^2$Physics Department, University of California Los Angeles,
Los Angeles, CA 90095--1547\\
$^3$Department of Physics, The Ohio State University,
Columbus, OH 43210\\
$^4$ Institute for Theoretical Physics, University
of California, Santa Barbara, CA 93106-4030\\
$^5$ Massachusetts Institute of Technology, 77 Massachusetts Ave,
Cambridge, MA 02139}

\date{\today}
\maketitle

\vskip 0.5 cm
\begin{abstract}
We discuss possible patterns of electron fractionalization in
strongly interacting electron systems. 
A popular possibility is one in which the charge of the electron 
has been liberated from its Fermi statistics. Such a fractionalized phase
contains in it the seed of superconductivity. Another possibility occurs when
the spin of the electron, rather than its charge, is
liberated from its Fermi statistics.
Such a phase contains in it the seed of magnetism,
rather than superconductivity.
We consider models in which both of these phases occur
and study possible phase transitions between them. 
We describe other fractionalized phases, distinct from these,
in which fractions of the electron  
themselves fractionalize, and discuss the topological
characterization of such phases. 
These ideas are illustrated with specific
models of $p$-wave superconductors,
Kondo lattices, and coexistence between $d$-wave
superconductivity and antiferromagnetism. 
 
\end{abstract}

\vspace{0.5 cm}

\vspace{1 cm}
]


\section{Introduction}

Electron fractionalization in strongly interacting electron 
systems in dimensions larger than one has been an important
subject of study since spin-charge separation
was suggested as a mechanism of high $T_c$
superconductivity \cite{Anderson87,Kivelson88} in the cuprates.
In particular, it was suggested that the electron is
splintered into a spin-carrying neutral excitation (`spinon')
and a charge-carrying spinless excitation 
(`holons' or `chargons'). There have been different proposals
in regard to this possibility, but the existence of such
phases in the cuprates is still controversial.  

On the other hand, there exist clear experimental examples of
phases in the quantum Hall regime of two 
dimensional electron systems where
quantum number fractionalization
has been well established.
The low energy excitations
(quasiparticles) in these two-dimensional strongly interacting 
electron systems carry fractions of the quantum numbers of 
the original electrons.
Different quantum Hall liquid states can be characterized
by different varieties of {\it topological order}. 
The transitions between different quantum Hall states can
be understood as topological-order-changing transitions which 
occur even in the absence of conventional broken symmetries.
The Hall conductance is but one of the topological quantum numbers
which characterize a given phase. Another
important property
of a topologically ordered state is the
ground state degeneracy of 
the system on higher genus manifolds such as tori. For each 
topologically ordered state, there are corresponding
sets of characteristic excitations with different quantum numbers.

It has become clear\cite{Wen1,topth} that the notion of
topological order also provides a 
precise characterization of spin-charge separated and other fractionalized 
phases in spatial dimensions higher than one even in situations of zero or 
weak magnetic fields. One of the remarkable features of the quantum
Hall effect is the 
enormously rich number of exotic phases which display 
different patterns of fractionalization of the electron and
associated topological orders. 
In view of the similarity between the theoretical characterization of 
quantum Hall states and fractionalized states in zero magnetic field,
it is tempting to investigate
a similar possibility of a variety of fractionalization patterns
in other strongly correlated systems. 
We explore this possibility in this paper. We describe
theoretically a few of the several different 
possible fractionalized phases that may exist in
various different models of strongly interacting electron systems. 

Following the introduction of the
Schwinger boson description of the
Heisenberg model of quantum antiferromagnets \cite{Arovas88},
slave fermion \cite{slave_fermion}
formulations of doped antiferromagnets were introduced.
In these formulations, it is assumed that the electron
decays into a bosonic, spin-$1/2$ {\it spinon} and a
fermionic, charge-$e$ {\it holon}. We will call this
phase {\it CfSb} ({\it C}harged {\it f}ermion,
{\it S}pinful {\it b}oson).

On the other hand, a phase with bosonic holons and
fermionic spinons -- which we will call
{\it CbSf} ({\it C}harged {\it b}oson,
{\it S}pinful {\it f}ermion) -- 
naturally leads to superconductivity
through the Bose condensation of bosonic holons 
in the presence of fermionic spinon pairing.
Consequently, much attention has been focused on the
description of such a fractionalized phase, especially
in the context of the slave boson description of the $t-J$ model.
The pairing symmetry of the resulting superconductor
is dictated by the
underlying symmetry of the spinon pairing. 

A $Z_2$ gauge theory of fermionic spinons
and bosonic holons was developed in the context of 
superconductivity in the cuprates \cite{Senthil99}
(see also \cite{RSSpN}, \cite{Jalabert91}, and \cite{Sachdev00f}).
Spinons and holons are coupled by 
an Ising gauge field. The deconfined phase of this theory
corresponds to the {\it CbSf} phase.
Most importantly, the deconfinement-confinement transition
of spinons and holons occurs through the condensation of 
vortices in the $Z_2$ gauge field, or {\it visons}. In the deconfined
phase, the visons exist as gapped excitations; when visons
condense, the spinons and holons are confined within 
electrons. The existence of gapped visons is crucial for
the robustness of the topological order of the deconfined fractionalized
phase \cite{Wen90,topth}.
Although this formalism was introduced in the
context of cuprate superconductivity, it
is sufficiently flexible to permit a
description of other types of fractionalized phases
including $CfSb$.

These ideas have a physical manifestation
in the context of quantum disordered magnets
and superconductors. In this picture, one visualizes
fractionalized states in terms of nearby ordered states.
In a broken (continuous) symmetry state, Goldstone modes
can screen the associated quantum number(s) \cite{Kivelson90}.
Thus it is possible for quasiparticles to
be stripped of some of their quantum numbers.
One might imagine that the destruction of
order by quantum fluctuations can preserve
this screening of quasiparticle quantum numbers.
This occurs when those topological defects of the ordered state
which braid non-trivially with the quasiparticles
persist as gapped
excitations even after the demise of the order \cite{Balents99}.
Indeed, this is precisely what happens when the
state is topologically-ordered. The neutral,
spin-$1/2$ fermion of the $CbSf$ state is viewed as
the descendent of the Bogoliubov-de Gennes quasiparticle;
the vison, of the $hc/2e$ vortex. When considered
in the context of spin-triplet superconductors and their
rich order-parameter structure, this immediately
suggests exotic phases such as $CbSf$, $CfSb$,
and even a third phase $CbSbNf$ ({\it C}harged {\it b}oson,
{\it S}pinful {\it b}oson,
{\it N}eutral {\it f}ermion), in which the charge-
and spin-carrying excitations are bosons and there is a neutral,
spinless fermionic excitation.
Since these superconductors can break both charge and spin
symmetries -- as do states in which singlet
superconductivity and magnetism coexist --
one can envision the screening of both
quantum numbers of a quasiparticle. If (the minimal)
topological defects in the charge sector
survive into a disordered state, then this
disordered state has neutral, spin-$1/2$ fermionic
excitations ($CbSf$); if topological defects in
the spin sector survive into a disordered state, then this
disordered state has charge-$e$ spinless fermionic
excitations ($CfSb$); if topological defects in
the {\it both} sectors survive into a disordered state, then this
disordered state has neutral, spinless fermionic
excitations ($CbSbNf$).

The analysis of quantum dimer models \cite{Kivelson88}
and resonating-valence-bond \cite{Anderson87,Anderson73,Kivelson87}
ground states led to conflicting claims that
the {\it CbSf} \cite{Kivelson87}  or {\it CfSb} \cite{Haldane89,Read89}
scenario is realized in these models.
These models have a
$Z_2$ vortex excitation \cite{Kivelson89,Haldane89}
-- which are precisely the {\it visons}
described above --
which are relative semions with spinons
and holons. Thus, a spinon or holon can
change between bosonic and fermionic
statistics by forming a bound state with a
vison. This begs the question whether
the {\it CbSf} phase discussed in
the context of superconductivity is the same as the {\it CfSb}
phase considered in relation to magnetism.
We reconsider this question in the context
of recent progress in the understanding of
fractionalized phases described above.
One might worry that the apparent differences between these
phases is an artifact of the formalisms
employed. One might also wonder if there
are any further fractionalized phases.
In this paper, we discuss the questions raised above
using three different models: 
p-wave superconductors, Kondo lattices, and
XY magnets coupled to d-wave superconductors.

The main results can be summarized as follows.

\vskip 0.2cm
\noindent
1. Both {\it CbSf} and {\it CfSb} phases can arise
in a variety of different models. 

\vskip 0.1cm
\noindent
2. Upon accepting the possibility of electron fractionalization,
one is led to consider a wide variety of fractionalized
phases. In the higher-level fractionalized phases, electrons
can be fractionalized in many different ways. For example,
spinons and holons can be further fractionalized.
Apart from the {\it CbSf} and {\it CfSb} phases,
we discuss two others. One is the {\it CSbNf}
({\it C}harge- and {\it S}pin-carrying {\it b}oson,
{\it N}eutral fermion) phase,
in which the electron breaks up into a boson which
carries both the spin and charge quantum numbers
and a neutral fermion. This phase is at the first level of
fractionalization along with the {\it CbSf} and {\it CfSb} phases.
The other is the {\it CbSbNf} phase,
in which there exist spin-carrying neutral bosons, charge-carrying
spinless bosons, and ``statistics-carrying'' neutral spinless
fermions. The {\it CbSbNf} phase is at the second level of
fractionalization. In principle, higher-level fractionalized phases exist.

\vskip 0.1cm
\noindent
3.  We demonstrate the existence of some of these
exotic phases in the context of the three different systems mentioned above
- Kondo lattices, $p$-wave superconductors, and models with both strong spin and
$d$-wave pairing fluctuations. 
For the $p$-wave superconductor, the four fractionalized
phases discussed here arise naturally  order parameter
has a rich spectrum of topological defects which
can condense in a variety of ways, thereby giving
rise to an array of fractionalized non-superconducting phases.

\vskip 0.1cm
\noindent
4.  The question of whether $CbSf$ and $CfSb$ are smoothly connected
to one another or whether they are necessarily 
separated by a phase transition is a subtle and delicate issue
for reasons that will be discussed at length later.
While we do not provide a definitive conclusion, we 
outline a possible scenario in which the distinction between
$CbSf$ and $CfSb$ is similar 
to that between liquid and gas phases.
These phases are separated by a first
order transition line which terminates at a critical point. 
In principle, one can go around the critical point from 
one phase to the other without encountering a phase transition.
This scenario is supported by a number of suggestive
(though certainly not conclusive)
arguments.

On the other hand, the transition between the
two phases can occur through another fractionalized phase
with a higher-level fractionalization pattern.
In this case, each transition in the process
could be a continuous transition.
We demonstrate
that the transition between {\it CbSf} phase and {\it CfSb}
phase can occur through the {\it CbSbNf} phase.

\vskip 0.1cm
\noindent
5. In order to examine whether one can go from {\it CbSf} to 
{\it CfSb} through further fractionalized phases like 
{\it CbSbNf}, 
one can design 
a {\it gedanken} flux trapping experiment similar to the one
proposed in \cite{toexp}. 
This {\it gedanken} experiment clearly demonstrates the existence
of a phase boundary between {\it CbSf} and {\it CfSb} 
when these phases are close to {\it CbSbNf}. 

\vskip 0.1cm
\noindent

\vskip 0.2cm
Topological order is robust against local perturbations
such as impurities.
Thus we will concentrate on general universal properties
of the fractionalized phases. One of our goals will
be to give a precise characterization of these phases
which is independent of the underlying microscopic models
where they may occur. We believe that these exotic phases
could play a role in the physics of $^3$He \cite{Leggett75}
and the ruthenates \cite{Maeno} as well as the cuprates
\cite{Bednorz},
organic superconductors \cite{organic},
heavy-fermion superconductors \cite{Stewart},
spinor Bose-Einstein condensates \cite{Zhou},
and the crusts of neutron stars \cite{NeutronStar}.

The rest of the paper is organized as follows.
In section II, we consider a Kondo lattice model
and how the $CfSb$ fractionalized phase can occur
in this model using the language of a $Z_2$ gauge theory.
Some details are given in Appendix A.
In section III, we suggest how this analysis
can be generalized and discuss a hierarchy
of fractionalized phases. Here we provide an overview of our
results.
%
%
In section IV, we discuss how this
hierarchy can be realized in $p$-wave superconducting
systems when the superconducting and spin order are
quantum disordered. This is done using the vortex
condensation formalism. In Appendix B,
the same ideas are shown to
apply to an XY magnet which is
coupled to a d-wave superconductor.
In section V, the fractionalized
phases of section IV are further discussed in the framework
of a $Z_2 \times Z_2$ gauge theory. 
In section VI, we consider the question
of the distinction in principle between
the putatively-different fractionalized
phases constructed in this paper.
In Appendix C, we give some technical details
of an argument using $Z_2 \times Z_2$ gauge theory
which supports our picture of the phase diagram.
In section VII, we show how flux-trapping
experiments (of the variety suggested by
Senthil and Fisher \cite{toexp})
can be used to shed further
light on the phase boundaries between these phases
and could be used to detect them.
We conclude in section VII.
Appendix D contains an aside in which we discuss
various novel properties of 
{\rm unfractionalized} phases occurring in the models
considered in this paper.

For other perspectives on fractionalization, see 
\cite{Anderson87,Kivelson88,Wen1,topth,Senthil99,RSSpN,Kivelson90,Balents99,Kivelson87,Kivelson89,Read89,toexp,z2short,Balents98,Balents00,RVB,Fradkin90,Sondhi01,Shtengel01}.

\section{Fractionalization in spin models: Spin-statistics separation}
\label{sss}

In principle, there are several possible ways in which the electron
can fractionalize in a strongly correlated system. In the context of the
cuprates, attention has focused on the situation in which the electron  
splinters into two separate excitations -a charged spinless boson, 
and a neutral spinful fermion. In this case, the charge of the electron 
is liberated from its Fermi statistics.  

In this section, we will briefly discuss another
possible fractionalization 
pattern in which the spin, rather than the charge,
of the electron is liberated from its Fermi statistics.
The electron splinters into a charged spinless fermion, and a 
spinful boson. As we will see, this phenomenon also requires the 
presence of a gapped topological $Z_2$ vortex excitation. 
The issue of whether such a fractionalized phase is {\em distinct}
from one in which the charge is liberated from the Fermi statistics
is a delicate one, and shall be discussed in the Section \ref{ddfp}.

To motivate the discussion, consider a ``Kondo-lattice" model with 
the Hamiltonian
\bea 
\label{kondo}
H & = & H_t + H_K + H_{ex} \\
H_t & = & - \sum_{<rr'>} t_{rr'} \left(c^{\dagger}_{r\alpha}c_{r'\alpha}
+ h.c \right) \\
H_K & = & J_K \sum_r \left(S^+_r c^{\dagger}_{r \da} c_{r\ua}
+ h.c \right) \\
H_{ex} & = & \sum_{rr'}-{J \over 2}\left(S^+_r S^-_{r'} + h.c \right) +
J_z S^z_r S^z_{r'}
\label{XXZ}
\eea
Here the $c_{i\alpha}$ represent ``conduction" electrons
with spin $\alpha$ at site $i$. 
The operators $\vec S_i$ are spin operators representing magnetic moments
localized at the lattice sites. The first term is the
usual conduction electron hopping,
described in a tight-binding approximation.
The second term is a ``Kondo"
coupling between the conduction electrons
and the local moments. The third term
is an explicit exchange interaction between
the local moments. For simplicity, 
we have assumed that system only has a $U(1)$
spin symmetry for rotations about the  $z$-axis of spin
(we will comment on situations with full $SU(2)$ spin symmetry later). 
We are interested not so much in establishing the exact phase diagram of this
particular model - rather our main interest here is in 
establishing the possible existence and stability
in models of this kind of quantum phases 
where the electron is fractionalized.
To that end, we will think more generally 
about a class of models which may be obtained from the model above by 
adding other local interactions which share its symmetries. 
If the system is in a quantum phase in which both the
symmetry of rotations about the 
$z$-direction of spin and the charge conservation symmetry is unbroken, 
the excitations may be labelled by 
their $S_z$ and charge ($Q$) quantum numbers.
Clearly, we can visualize two qualitatively
different possibilities. First, the system may be in a phase in which the 
excitations are electrons ($Q = 1, S_z = \frac{1}{2}$)
or composite objects made from 
electrons (such as, for instance, a magnon which has $Q = 0, S_z = 1$). 
This is a 
conventional phase of the kind familiar from textbooks
(for instance, a Fermi liquid or a 
band insulator). On the other hand, one could also imagine phases
in which there are excitations which carry quantum numbers 
which are fractions of those of an electron. The simplest possibility 
(the one we will focus on)
is that there are excitations which carry $S_z = 1/2, Q = 0$ (``spinons")
and others which carry 
$S_z = 0, Q = 1$ (``holons"). In such a phase, 
the electron has been fractionalized. In what follows,
we will discuss several ways of thinking
about such phases. Our focus will be on general universal
properties of such phases. 
In particular, we will be interested in obtaining robust
precise characterizations of 
fractionalized phases that are independent of the particular
microscopic models in which
they possibly occur.

It is extremely instructive to begin by just considering the physics of the 
local moments alone as described by the exchange part
of the Hamiltonian $H_{ex}$.
This Hamiltonian is clearly invariant under a global
spin rotation about the $z$-axis
of spin. For technical simplicity, we will assume $J, J_z \geq 0$. 
The physics of this particular Hamiltonian is well-understood:
when $J_z/J$ is small,
there is long range order in $S^+$. When $J_z/J$ is large, the 
system breaks translational symmetry with $<S^z>$ being larger in one 
sublattice of the square lattice
than the other, but
the $U(1)$ spin rotation symmetry is unbroken. 
The point $J_z = J$ can be mapped to the nearest-neighbour
antiferromagnetic Heisenberg model with 
full $SU(2)$ spin symmetry on a bipartite lattice by
rotating the spins on one sublattice 
by $\pi$ about the $z$-axis.  In the specific case of
a square lattice (which we assume through out 
our discussion), this   
is known to develop Neel long range order in
two spatial dimensions. Our interest here is not so
much in the properties of this
{\em particular} Hamiltonian as in the properties of
an entire {\em class} of systems
with the same symmetry, and with short-ranged
interactions between the spins. 
In particular, we will
be interested in fractionalized phases in which the
excitations are spinons with 
quantum number $S_z = 1/2$.
To that end, we will reformulate the Hamiltonian
directly in terms of ``spinon'' fields
which carry spin $S^z = 1/2$. 
This naturally introduces a $Z_2$ gauge symmetry. 
The result is a theory of bosonic spinon
fields coupled to a $Z_2$ gauge field
which can then be used to analyze the
possibility of fractionalized phases and 
their universal properties.

We may think of $S^+, S^-$ as the creation and destruction
operators respectively
of a hard core boson on the sites of the lattice. Specifically, write 
$S^+_{r} \equiv b^{\dagger}_{sr}$, $S^-_{r'} \equiv b_{sr}$, and 
$S^z_r = 1/2 - b^{\dagger}_{sr} b_{sr}$. Note that there is half a
boson for each site on average. Now imagine relaxing the hard-core 
constraint on the bosons, and instead add a term
\be
\frac{U}{2} \sum_r (2n_r - 1)^2
\ee
at each lattice site. Here $n_r$ is the boson number at each site. 
In the limit $U \ra \infty$, we  recover the spin model exactly. 
For large but finite $U$
however, relaxing the hard-core constraint is expected to be innocuous. 
It is now convenient to go to a number-phase representation for the 
bosons: we write $b_{sr} \sim e^{i\varphi_r}$ with 
$[\varphi_r, n_{r'}] = i\delta_{rr'}$. For simplicity, we also 
specialize to the limit where $J_z = 0$. The Hamiltonian then becomes
\be
\label{hexbos}
H = \sum_{<rr'>} - J\, \cos(\varphi_r - \varphi_{r'}) +
\frac{U}{2}\sum_r (1 - 2n_r)^2
\ee
This is clearly closely related to the original 
spin Hamiltonian in Eq.\ref{XXZ}. 
Now consider a formal {\em change of variables} which
involves splitting the boson operator
$b_{sr}$ into two pieces:
\bea
\label{splitsp}
b_{sr} & =  e^{i\varphi_r} &  =  z^2_{r} \\
z_r & \equiv  e^{i\phi_r} & =  s_r e^{i\frac{\varphi_r}{2}}~~ (s_r = \pm 1)
\eea
We will refer to $z_r$ as the spinon destruction operator.
Note that with these definitions, both $\varphi_r$ and
$\phi_r$ are defined in the interval
$[0, 2\pi)$. It is also convenient to define a number
operator for the spinons $N_r = 2n_r$ 
which is conjugate to $\phi_r$. 
In terms of the spinon operator, the Hamiltonian becomes
\be
\label{hexsp}
H = \sum_{<rr'>} -J \,\cos(2\phi_r - 2\phi_{r'}) +
\frac{U}{2}\sum_r (N_r - 1)^2 .
\ee
The change of variables above must be supplemented with a
constraint - clearly the 
physical Hilbert space consists only of states where
$N_r$ is even. Therefore, 
we need to impose the operator constraint
$(-1)^{N_r} = 1$ at each site of the lattice. 
Formally this may be implemented through the projection operator
\bea
{\cal P} & = & \prod_r {\cal P}_r \\
{\cal P}_r & = & \frac{1}{2} \left(1 + (-1)^{N_r} \right)
\eea
Note that $[{\cal P}, H] = 0$. It is now
convenient to pass to a functional integral formulation. 
We follow Ref. \cite{Senthil99,z2short} closely to obtain for the partition 
function
\bea 
Z & = & \sum_{\sigma_{r\tau}} \int {\cal D}\phi e^{-S} \\
S & = & S_{\tau} + S_{r} + S_B \\
S_{\tau} & = & \sum_{\tau,r} J_{\tau} \sigma_{r\tau}
\,\cos(\phi_{r, \tau + \epsilon}
- \phi_{r\tau}) \\
S_{r} & = & \epsilon \sum_{<rr'> \tau} J \,\cos(2\phi_{r\tau} - 2\phi_{r'\tau})
\eea
where $\sigma_{r\tau} = \pm 1$ may be interpreted
as the time-component of a $Z_2$
gauge field that imposes the constraint on the Hilbert space, and $\epsilon$
is the lattice spacing along the time direction. The constant $J_{\tau}$ is  
determined by the original interaction strength $U$. 
The term in the action $S_r$ involving the spatial coupling may be decoupled
by a Hubbard-Stratanovich transformation:
\be
e^{-S_r} = \int D\chi e^{-\epsilon J \sum_{<rr'>, \tau} \chi_{rr'}(\tau)^2 
+ 2\epsilon J \chi_{rr'}(\tau) \left(z^{\dagger}_{r}(\tau) z_{r'}(\tau)
+ c.c \right)}
\ee
Here $\chi_{rr'}(\tau)$ is a real-valued field.
We have omitted an unimportant overall
constant.  

We now proceed exactly as in Ref. \cite{Senthil99,z2short},
and replace the integral over the 
{\em continuous} variable $\chi$ by a sum over a
discrete field $\sigma_{rr'}(\tau) = \pm 1$. 
As discussed in Ref. \cite{Senthil99,z2short}, this approximation respects
all the symmetries of the action, 
and is expected to be innocuous. The resulting partition function becomes
\bea
Z & = & \sum_{\sigma_{ij}} \int {\cal D}\phi e^{-S} \\
S & = & S_s + S_B \\
S_s & = & -\sum_{<ij>} J_{ij} \sigma_{ij} \,\cos (\phi_i - \phi_j)
\eea
Here the $i,j$ label the sites of a space-time
lattice in three dimensions. The constants
$J_{ij}$ = $J_{\tau}$ for temporal links, and equals $\epsilon J$ for
spatial links. $S_B$ is the Berry phase action
\begin{eqnarray}
S_B = \frac{\pi i}{2} \sum_{i,j=i+\hat{\tau}} ( 1 - \sigma_{ij}) 
\end{eqnarray}
Note that the action (18)-(20) respects all the symmetries of the 
original model. The discrete field $\sigma_{ij} = \pm1$ may be viewed as a 
$Z_2$ gauge field. At this stage, this field does not have any dynamics. 
However, it is natural to expect that upon coarse-graining, some dynamics will
be generated. The simplest such term allowed by symmetry is
\be 
S_K = -K \sum_{\Box} \prod_{\Box} \sigma_{ij} . 
\ee
We will therefore consider the full action 
\be
S = S_s + S_K + S_B
\ee

What we have achieved so far is an approximate reformulation of  
spin models with $XXZ$ symmetry. This reformulation is extremely useful to 
explore the various possible allowed phases in such models. However, the 
approximations made in obtaining this reformulation are severe enough that 
it is not easy to see which one of these allowed phases will
be obtained in any particular microscopic model.  

Consider the possible phases when the parameter
$K$ is very large. When $K = \infty$, 
the $Z_2$ flux through each plaquette is constrained to be one.
We may then choose a gauge in
which $\sigma_{ij} = 1$ on every link. In this limit therefore,
the action reduces to
\be
S =  -\sum_{<ij>} J_{ij}  \,\cos (\phi_i - \phi_j) 
\ee
This simply describes a quantum $XY$ model in two spatial dimensions.
Note that the 
Berry phase term simply vanishes when all the $\sigma_{ij} = 1$. There 
clearly are two possible phases - an $XY$ ordered phase in which
$z_{i} = e^{i\phi_i}$
has condensed, and a paramagnetic phase in which the excitations
created by $z_{i}$
are gapped. Note that these excitations in the paramagnetic phase 
carry spin $S^z = 1/2$. Thus, the spin has been fractionalized in this phase. 

Now consider moving away from the limit $K = \infty$ by
making $K$ large but finite.
For finite $K$, as can be seen from the arguments
advanced in Ref. \cite{Senthil99}, 
the $XY$ ordered phase where the spinon field has
condensed is indistinguishable from 
a conventional $XY$ ordered $XXZ$ magnet. The paramagnetic
phase in which the spinons are 
uncondensed and deconfined survives for large but finite $K$.
When $K$ is finite, 
it becomes clear that this phase has another distinct
excitation which carries the flux of 
the $Z_2$ gauge field. This $Z_2$ vortex - dubbed the vison -
does not carry any 
physical spin, and has an energy gap 
of order $K$ for large $K$. It has the important
property that when a spinon is taken around it, 
the wavefunction of the system acquires a phase of $\pi$. 
  
Upon decreasing $K$, at some critical value,
the vison gap goes to zero. For smaller $K$,
the visons condense leading to confinement of the spinons.
The resulting phase is a 
conventional quantum paramagnet with gapped $S_z = 1$ excitations.
In this phase, 
the Berry phase term becomes important and leads to a breaking of
translational symmetry
- the paramagnet is therefore expected to develop spin-Peierls order.
We will not discuss
such conventional phases very much in this paper. 

Much further insight into the physics of the fractionalized phases
may be obtained
by the following considerations.
We begin by first considering ordered phases in which the
symmetry of rotations about the 
$z$ direction of spin has been broken spontaneously.
For simplicity, we consider 
a phase in which the spins have all lined up along some
direction in the $xy$ plane. 
The general properties of such a phase are well-known.
There are two distinct 
kinds of excitations. First, there is a gapless spin-wave
mode with linear dispersion. 
Apart from these, there are also topological vortex excitations.
On moving along any circuit
that encloses a vortex, the direction of the spin in the
$xy$ plane winds by an integer 
multiple of $2\pi$. This integer winding number - the
vorticity - is conserved, and may be used to 
label the spectrum of excited states. States with different total vorticity 
belong to different topological sectors and are not mixed by the dynamics
generated by the Hamiltonian.
Note that in this ordered phase we can no longer label states by their
$S^z$ quantum number. 

These familiar properties of the $XY$ ordered phase must be
contrasted with 
those of the quantum paramagnet. First consider a
conventional paramagnet ({\em i.e.}
one with no fractionalization). Clearly in this phase $S^z$ is conserved, and 
is a good quantum number to label the excitation spectrum.
On the other hand, the vorticity
loses its meaning in the paramagnetic phase, and is
no longer a good quantum number. 
This suggests that one may view the paramagnet as a phase in which the vortex 
excitations have themselves condensed. Condensation of the vortices implies 
that the vorticity is no longer a good quantum number
(just like condensation of 
spin implies that $S^z$ is no longer a good quantum number). Indeed, these 
observations may be formalized precisely by means of a duality transformation
which reformulates the system in terms of the vortex
fields rather than the spins. 
In this dual formulation, the paramagnet is described
as a vortex condensate, and the
$XY$-ordered phase as a vortex insulator (in which the vortices are gapped). 
The physical excitations of the paramagnet which carry
the $S^z$ quantum number 
appear as dual flux tubes of the vortex condensate in this language.  

How are we to view the fractionalized quantum
paramagnet in this dual language? 
As the phase in question is a paramagnet, it is clear that the vorticity
has no meaning, implying that the vortices must have condensed.
As pointed out in Ref. \cite{Balents99},
we may view the fractionalized phase as a condensate of {\em paired} vortices. 
This has the immediate consequence of halving the dual flux tube, {\em i.e} of 
fractionalizing $S^z$ as required. Furthermore, note
that the unpaired (and uncondensed)
single vortex is still an excitation in the system. Its vorticity is screened 
by the (double strength) vortex condensate as is required in the
paramagnet. However, 
its parity is still a good quantum number.
Thus the unpaired vortex, though a legitimate
excitation of the fractionalized paramagnet, carries only a 
$Z_2$ quantum number - it is clear that it is the vison excitation discussed
previously.

The discussion above provides a description of a
fractionalized quantum paramagnet
in the context of spin models with $XXZ$ symmetry. 
We now return to the full model which includes coupling to the 
``conduction" electron degrees of freedom. As above, we first replace the 
operator $S^-_r$ in the Kondo coupling at each site by the boson operators
$b_{sr} \sim e^{i\varphi_r}$ (and similarly for $S^+_r$). The Kondo coupling
term then becomes
\bea
\label{kb}
H_K & = & J_K \sum_r \left (b^{\dagger}_{sr}
c^{\dagger}_{r\da}c_{r\ua} + h.c. \right) \\
& = & J_K \sum_r \left (z^{2\dagger}_{r} c^{\dagger}_{r\da}c_{r\ua}
+ h.c. \right)
\label{ks} 
\eea
In going to the second equation, we have introduced the
spinon operators $z_r$ defined in (\ref{splitsp}). 
The Kondo coupling can be further simplified by another change of variables 
\bea
\label{splitcua}
\eta_{r\ua} & \equiv & z_rc_{r\ua} \\
\eta_{r\da} & \equiv & z^{\dagger}_{r}c_{r\da}
\label{splitcda}
\eea
We will call the $\eta$-operators the holon operators.
In terms of the holons, the Kondo
coupling becomes
\be
H_K = J_K \sum_r \left (\eta^{\dagger}_{r\ua} \eta_{r\da} + h.c. \right)
\ee
Note that the holons are actually {\em spinless} charge $e$ fields
despite the presence of the label $\ua, \da$. This is obvious from their
definition in terms of the spinon and electron operators above: the holon
operators do not transform under spin rotations about the $z$-axis. 
Explicitly, the 
Kondo term mixes up and down holons so that their label $\ua, \da$ is
changed by the dynamics. Therefore their spin label has no 
great significance, and they are correctly viewed as spinless fermions.
We may use the following physical picture: the Kondo
spins screen the spin of the conduction electrons. 

Under these changes of variables, the electron hopping term becomes
\be
\label{htcs}
H_t = -\sum_{<rr'>} t_{rr'} \left[ z^{\dagger}_{r} z_{r'}
\left(\eta^{\dagger}_{r'\ua}
\eta_{r'\ua} + \eta^{\dagger}_{r\da} \eta_{r'\da} \right) + h.c. \right]
\ee
We now make approximations very similar to
those used above for the exchange part of the 
Hamiltonian. They allow us to reformulate the system in terms of the
spinons, holons and a $Z_2$
gauge field. Some of the details are oulined in the Appendix. 
The resulting action can essentially be guessed on
symmetry grounds, and takes the form:
\bea
\label{z2kon}
S & = & S_c + S_s + S_B + S_K \\
S_c & = & - \sum_{<ij>} 
\sigma_{ij} t^c_{ij}(\eta^{\dagger}_{i\ua} \eta_{j\ua} +
\eta^{\dagger}_{i\da} \eta_{j\da} + c.c. ) \\ \nonumber
& + & J_{K} \sum_i (\eta^{\dagger}_{i\ua} \eta_{j\da} + c.c.)
\eea
The other terms of the action are as given before. 
 
Following the discussion above, for large $K$, we expect to
have a phase in which 
the holons and spinons are liberated from each other.
In such a phase, the 
electron is fractionalized. However, in contrast to the
fractionalized phase that is most 
popular in the context of the cuprates, here the spin of the electron has
been liberated from its Fermi statistics.
Are these two phases actually the same?
We will address this issue in subsequent sections.

Though we have based our discussion on models with $XXZ$ symmetry,
we expect the  fractionalized quantum
paramagnetic phase to exist even in systems with 
full $SU(2)$ spin symmetry. Indeed, in the context of
frustrated quantum $Sp(n)$ spin models in the large-$n$ limit,
Read and Sachdev\cite{RSSpN} have argued for the
stability of fractionalized paramagnetic phases
with properties similar to that discussed above.

\section{A Hierarchy of Fractionalized Phases}

In Section \ref{sss}, we primarily discussed
fractionalized phases in which the 
electron splinters into a spin-$1/2$ neutral boson
and a charged spinless fermion. 
For future convenience, we will refer to this as
the $CfSb$ (charged fermion, spinful boson)
phase.
This is to be contrasted with the fractionalized
phases which are popular in 
the context of cuprate physics in which the electron
splinters into a spin-$1/2$ 
neutral {\em fermion} and a charged spinless {\em boson}
(see also Sections IV and V). 
We will refer to this as the $CbSf$ phase.  
In both cases, there is, in addition, a 
$Z_2$ vortex excitation (the vison) such that
taking either the holon or spinon
around it produces a phase change of $\pi$.

\begin{figure*}[h]
\epsfysize=10.0cm 
\epsfbox{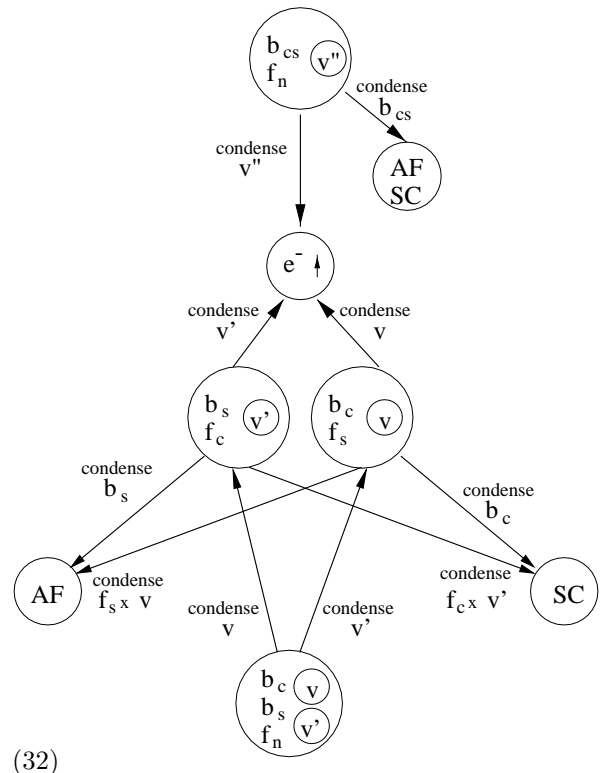}
\caption{Hierarchy of fractionalized phases}
\label{sfig}
\end{figure*}

Having accepted the possibility of quantum number fractionalization,
one can imagine
a wide variety of possible phases apart from the
two mentioned above. In particular, 
one may consider exotic possibilities where the fractions of the electron in 
any given fractionalized phase themselves fractionalize.
Such phases may be considered  to have a  
higher level of fractionalization. To see how
these may be described in the same 
kind of formulation
as discussed in this section, consider the following action:
\bea
\label{eqn:big-action}
S & = & S_{f} + S_c + S_s + S_{\sigma\tau} \, \cr  
S_{f} &=& - \sum_{ij, \alpha} \sigma_{ij} \tau_{ij}
[ t^n_{ij} {\bar f}_{i \alpha} f_{j \alpha} 
+ {\tilde t}^{\Delta} a_{ij}( 
f_{i\ua} f_{j\da} - (\ua \ra \da)) + c.c. ]\cr
& & -  \sum_{i \alpha} {\bar f}_{i\alpha} f_{i\alpha} \ , \cr
S_c &=& - \sum_{ij} t^c_{ij} \tau_{ij} 
(b^*_{ci} b_{cj} + c.c.) \ , \cr
S_s &=& - \sum_{ij} t^s_{ij} \sigma_{ij} 
(z_{i}^* z_{j} + c.c. ) 
\ , \cr
S_{\sigma\tau} & = & -K_{\sigma} \prod_{\Box}
\sigma_{ij} - K_{\tau} \prod_{\Box} \tau_{ij}
- K_{\sigma\tau} \prod_{\Box} \sigma_{ij} \tau_{ij}.
\eea
Here $b_c$ is a charge $e$ spinless boson and $z$
is a spin-$1/2$ chargeless boson. 
The $f$ field represents a spinless, 
neutral fermion (the spin index is just a 
label with no special significance).
The $\sigma_{ij}$ and $\tau_{ij}$ are two independent
$Z_2$ gauge fields. The physical electron
$c_{i\alpha} = b_{ci}z_{i} f_{i\alpha}$. 
Clearly if the $\tau$ field is confining,
the $f_{\alpha}$ and $b_c$ get confined to form a fermionic
holon - we then recover the action discussed
earlier in this section. On the other hand, 
if the $\sigma$ field is confining, the Fermi
statistics gets glued to the spinon ($z_{i}$)
- the resulting theory is essentially that introduced
in Ref. \cite{Senthil99} in the context of 
cuprate physics and involves {\em bosonic}
holons and {\em fermionic} spinons coupled to a 
$Z_2$ gauge field. If both gauge fields $\sigma$
and $\tau$ are deconfining however, we have an exotic
phase in which the fields $b_c, z, f$ are all liberated. 
This phase will also have two distinct vison
excitations corresponding to the fluxes of the two $Z_2$ gauge fields. 
We may view this phase as a
higher-level fractionalized phase as compared to the one discussed in Ref. 
\cite{Senthil99} or that
discussed earlier in this section. The
connection between various fractionalized phases 
is shown in Figure \ref{sfig}.
We use symbols $b_a$ and $f_a$ to label bosons and fermions that carry
quantum numbers $a=n,c,s,cs$ (neutral, charge, spin, charge and spin) and 
show the existence of appropriate $Z_2$ vortices in each phase
(for more details see Section V).

In the sections which follow, we will show how
an effective action such as that of Eq. \ref{eqn:big-action}
can arise in the context of $p$-wave superconducting systems
and systems which feature interplay between
magnetism and superconductivity.

\section{Fractionalization of electron quantum
numbers with $p$-wave pairing}

\subsection{Order Parameters and Symmetries}

Spin-triplet superconductors and their
rich order-parameter structure offer
the prospect of various exotic phases.
Since they break both charge and spin
symmetries, triplet superconductors
exhibit features of both singlet
superconductors and of spin models.
In particular, we can envision the restoration
of the the ${U_C}(1)$ charge symmetry by quantum fluctuations,
thereby resulting in a spin-triplet insulating state.
Alternatively, the spin symmetry
(we will make the simplifying assumption
that the system has only an easy-plane
${U_S}(1)$ spin symmetry)
can be restored,
resulting in a spin-singlet superconducting state.
Finally, both symmetries can be restored, leading
to a singlet insulating state. We believe that the
gapped, symmetry-restored states will not be very sensitive
to the precise symmetry of the spin sector,
so we believe that our results apply to systems with
full $SU(2)$ spin symmetry as well. In particular, when
the symmetry is increased (while keeping the size of the
representation fixed), fluctuations are enhanced,
and a system is more likely to be in a disordered
state. In order for these symmetries to be
restored separately, it will be necessary,
as we discuss below, for a type of topological ordering
to occur. This topological ordering is essentially
spin-charge separation of the charge $2e$, spin-triplet
Cooper pairs.
Depending on the way in which the symmetries
are restored, it is possible for further
topological ordering to take place, in which
case the quantum disordered
states may support excitations with exotic
quantum numbers. In such states, the spin and/or
charge of the quasiparticles is screened
by the Goldstone modes (which are themselves
separated from each other by the higher-level
topological ordering). As we describe in
this paper, there are no fewer than nine phase
which can result in this way.


To be concrete, let us consider the following
$p$-wave superconducting state of electrons on a square
lattice:
\begin{equation}
\label{eqn:triplet_op_def}
{\Delta_{\alpha\beta}} = {\Delta_0}\,\,\,{e^{i\varphi}}
\,\left(\cos\theta\,{\sigma^z_{\alpha\beta}}
\,+\,i\sin\theta\,{\delta_{\alpha\beta}}\right)\,\,\sin{k_y}a
\end{equation}
This is the most general unitary triplet state
in $2D$ \cite{Leggett75} if we assume that there is
only the $U(1)$ spin symmetry of rotations about
the $z$-axis, rather than the full $SU(2)$.
In (\ref{eqn:triplet_op_def}), only
${\Delta_{\uparrow\uparrow}}$ and
${\Delta_{\downarrow\downarrow}}$ are non-zero.
The lower symmetry could be the result of
spin-orbit coupling.
The symmetry-breaking
pattern associated with this order parameter is:
${U_C}(1)\times {U_S}(1)\times {D_4}\rightarrow {Z_2}\times
{Z_2}\times {D_2}$. The ${U_C}(1)$ charge symmetry is broken
to $Z_2$ by the condensation of a charge $2e$ order parameter.
The ${U_S}(1)$ spin-rotational symmetry is completely broken.
The square lattice point group, $D_4$, is broken to $D_2$ by the orbital
symmetry of $\Delta$. Finally, there is an additional
$Z_2$ since the order parameter is left invariant
by  $\varphi \rightarrow \varphi + \pi$, 
$\theta \rightarrow \theta + \pi$. As we discuss later,
this can be understood as a $Z_2$ gauge symmetry. From
${e^{i\varphi}}$
and ${e^{i\theta}}$ we can construct the following
$Z_2$-invariant order parameters whose
presence or absence characterizes the
phases which we consider.
In the absence of the triplet $p$-wave
superconducting order parameter (\ref{eqn:triplet_op_def}),
we can characterize states by the
charge-$4e$ order parameter,
\begin{equation}
{\Delta^{4e}} = \left({e^{i\varphi}}\right)^2
\end{equation}
and the spin nematic order parameter,
\begin{equation}
{Q} = \cos 2\theta
\label{Qop}
\end{equation}

These order parameters define the following
quantum-disordered states of triplet $p$-wave
superconductors.
\begin{itemize}
\item Charge-$4e$ singlet superconductor:
${\Delta^{4e}}\neq 0$, ${Q} = 0$.

\item Charge-$4e$ nematic superconductor:
${\Delta^{4e}}\neq 0$, ${Q} \neq 0$.

\item Spin-nematic insulator: ${\Delta^{4e}}= 0$,
${Q} \neq 0$.

\item Spin-singlet insulator: ${\Delta^{4e}}= 0$,
${Q} = 0$.
\end{itemize}

\subsection {Topological Defects}
\label{td}

The quantum-disordered and topologically-ordered
states which we will consider
can be understood in terms of the condensation
or suppression of various topological
excitations. The most basic and fundamental
topological excitation
is a composite formed of
a flux $hc/4e$ vortex together
with a $\pi$ disclination \cite{Vollhardt}.
\begin{figure*}[h]
\epsfxsize=7.0cm 
\epsfbox{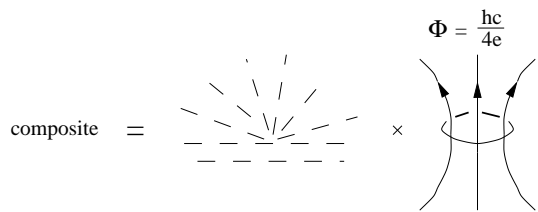}
\caption{$\pi$-disclination - $hc/4e$ vortex composite}
\label{composite}
\end{figure*}
Along a circuit about such an excitation, both $\varphi$ and
$\theta$ wind by $\pi$ so that any
$Z_2$-invariant combination is single-valued.
If such an excitation is at the origin, and $r$, $\phi$
are polar coordinates in the plane, then the
order parameter is of the form:
\begin{eqnarray}
{\Delta_{\alpha\beta}}(r,\phi) &=&
{\Delta}(r)\,\,\,{e^{i{\epsilon_1}\phi/2}}
\,\,\times\cr
& &
\,\left(\cos\frac{\phi}{2}\,{\sigma^z_{\alpha\beta}}
\,+\,i{\epsilon_2}\sin\frac{\phi}{2}\,
{\delta_{\alpha\beta}}\right)\,\,\sin{k_y}a
\end{eqnarray}
where ${\Delta}(0)=0$ and ${\Delta}(\infty)={\Delta_0}$.
The flux is into or out of the plane, respectively,
for ${\epsilon_1}=\pm 1$; the spins wind clockwise
or counter-clockwise, respectively,
for ${\epsilon_2}=\pm 1$.
It is instructive to write this as:
\begin{eqnarray}
{\Delta_{\uparrow\uparrow}}(r,\phi) &=&
{\Delta}(r)\,\,\,{e^{i {\epsilon_+}\,\phi}}
\,\,\sin{k_y}a\cr
{\Delta_{\downarrow\downarrow}}(r,\phi) &=&
-{\Delta}(r)\,\,\,{e^{i {\epsilon_-}\,\phi}}
\,\,\sin{k_y}a
\end{eqnarray}
where ${\epsilon_\pm}=({\epsilon_1}\pm{\epsilon_2})/2$.
Hence, $\pi$-disclination - $hc/4e$ vortex composites
are vortices in ${\Delta_{\uparrow\uparrow}}$ or
${\Delta_{\downarrow\downarrow}}$ alone.

These excitations can be combined to form an
$hc/2e$ vortex which is non-trivial in the
charge sector but trivial in the spin sector.
\begin{eqnarray}
{\Delta_{\alpha\beta}}(r,\phi) = {\Delta}(r)\,\,\,{e^{i\phi}}
\,\,\left(\cos{\theta_0}\,{\sigma^z_{\alpha\beta}}
\,+\,i\sin{\theta_0}\,{\delta_{\alpha\beta}}\right)\,\,\sin{k_y}a
\end{eqnarray}
with constant $\theta_0$.
Alternatively, we can form merons, which are
trivial in the charge sector but not the
spin sector.
\begin{eqnarray}
{\Delta_{\alpha\beta}}(r,\phi) = {\Delta}(r)\,\,\,{e^{i\varphi_0}}
\,\,\left(\cos\phi\,{\sigma^z_{\alpha\beta}}
\,+\,i\sin\phi\,{\delta_{\alpha\beta}}\right)\,\,\sin{k_y}a
\end{eqnarray}
with constant $\varphi_0$.
Finally, there are various
composites formed from the above.
A composite formed by $n$ $hc/2e$
vortices and $m$ merons takes the form
\begin{eqnarray}
{\Delta_{\alpha\beta}}(r\rightarrow\infty,\phi) =
{\Delta_0}\,\,\,{e^{i n\phi}}
\,\,\left(\cos m\phi\,{\sigma^z_{\alpha\beta}}
\,+\,i\sin m\phi\,{\delta_{\alpha\beta}}\right)\,\,\sin{k_y}a
\end{eqnarray}

\begin{figure*}[h]
\centerline{\epsfxsize=9.0cm 
\epsfbox{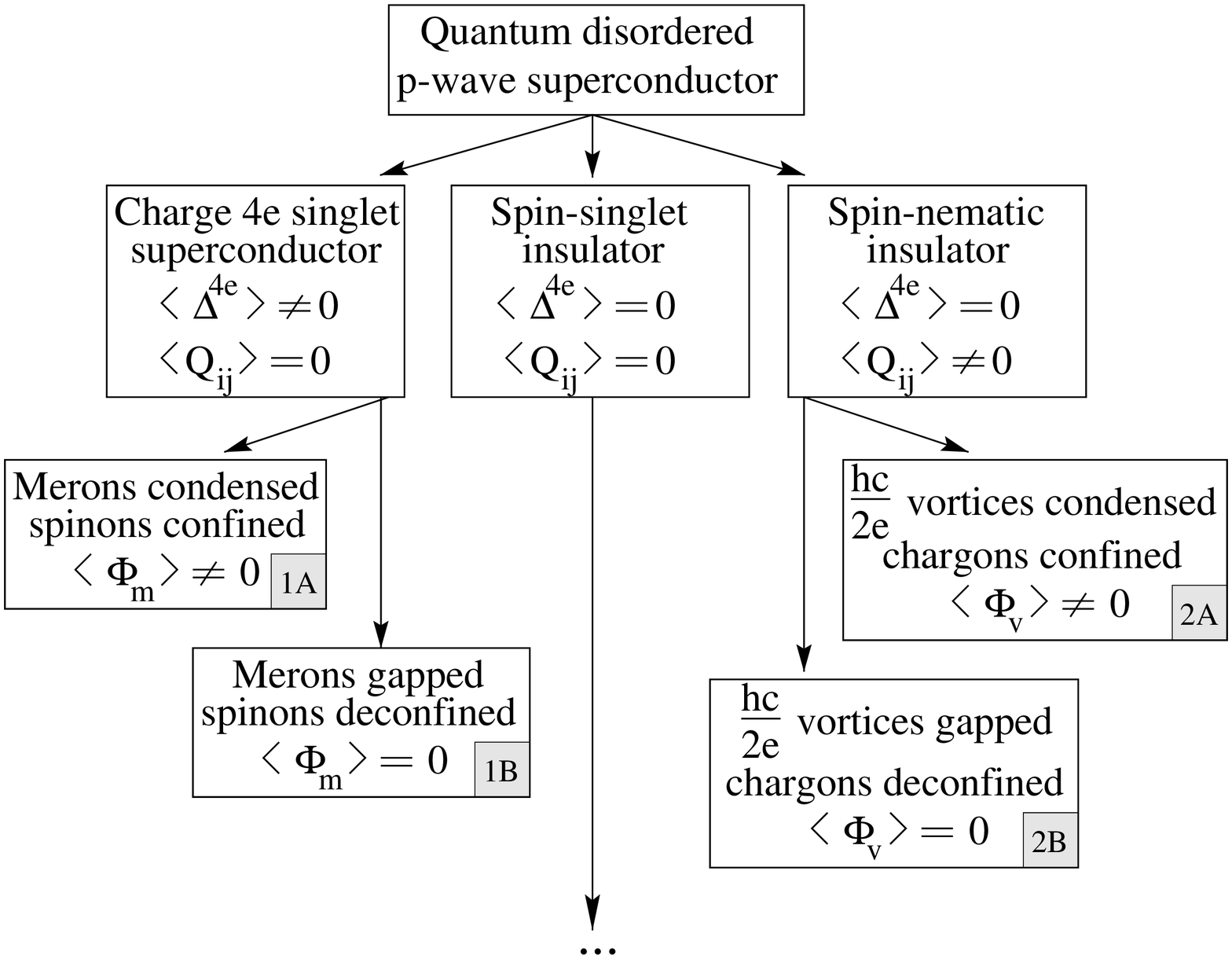}}
\caption{Phases of quantum-disordered $p$-wave superconductor.}
\label{phase1}
\end{figure*}

\begin{figure*}[h]
\centerline{\epsfxsize=9.0cm 
\epsfbox{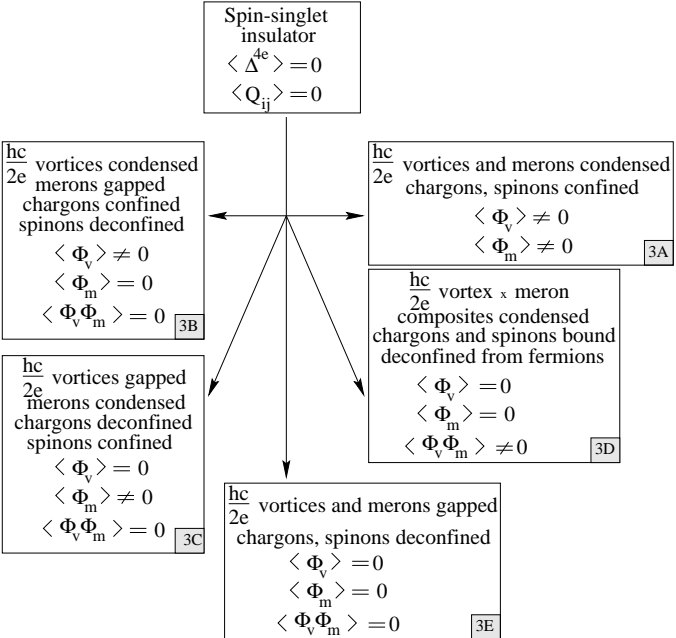}}
\caption{Phases of quantum-disordered $p$-wave superconductor.
Note that the phase in which $hc/2e$ vortices and merons are condensed may
be described as having $hc/4e$-vortex-$\pi$ composites condensed.}
\label{phase2}
\end{figure*}

If flux $hc/4e$ vortex-$\pi$ disclination
composites condense, then ${U_C}(1)$ and ${U_S}(1)$
are restored. The system will be in a
singlet insulating state and all excitations
will have conventional quantum numbers.
If, on the other hand, $hc/4e$ vortex-$\pi$ disclination
composites are gapped and only complexes
consisting of multiples
of $hc/4e$ vortex-$\pi$ disclinations (e.g.
$n$ $hc/2e$-$m$ meron composites)
are condensed, then quantum number separation is possible.
If complexes consisting of a multiple of
four $hc/4e$ vortex-$\pi$ disclinations condense,
then we will have the various versions of quantum number separation
summarized in Figures \ref{phase1} and \ref{phase2}.

\subsection{Quantum Number Separation}
\label{qns}

The effective action of a $p$-wave superconductor
may be written in the form:
\begin{equation}
\label{eqn:S_tot}
{S_{\rm tot}} = {S_f} + {S_c} + {S_\sigma}
\end{equation}
where ${S_f}$ is the action for the fermionic
quasiparticles and their interactions with
the Goldstone modes, and ${S_c}$ and ${S_\sigma}$
are the actions for the charge and spin Goldstone
modes.

\begin{figure*}[tbh]
\centerline{\epsfxsize=6.0cm 
\epsfbox{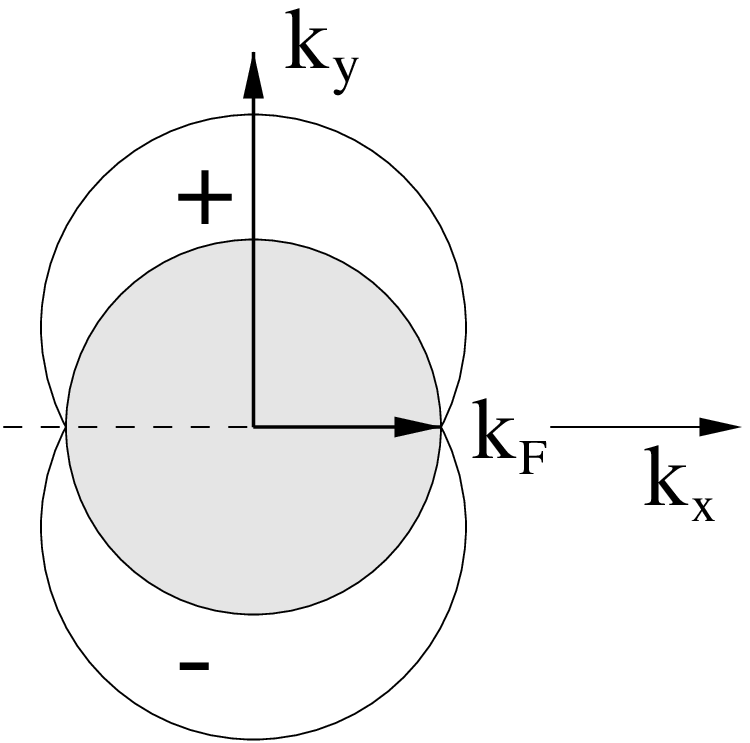}}
\caption{ Order parameter for a $sin k_y$ $p$-wave superconductor.
Gapless excitations exist at $\vec{ k_F} = (\pm k_F,0)$.}
\label{pnodal}
\end{figure*}

Depending on the topology of
the Fermi surface, the low-energy spectrum of a $p$-wave superconductor may 
include gapless fermionic quasiparticles. Let us assume that
the topology is such that the gap has nodes on the Fermi surface.
Focusing on the nodes, as shown in figure \ref{pnodal}.
We linearize the action:
\begin{eqnarray}
\label{eqn:Gold-qp-int1}
{S_f} &=&  \int {d^2}x\,d\tau\,\chi^\dagger
\biggl[ \partial_\tau - {A^c_\tau}{\tau^z}
 - {v_F} {\tau^z}
  i\partial_x + {v_F} {A^c_x}\cr
& & {\hskip 1.9 cm}-\, A^{\sigma}_{\tau} \sigma_z + 
v_F A^{\sigma}_x \sigma_z \tau_z\cr
& &  -  v_{\scriptscriptstyle\Delta}
\tau^s   e^{is\varphi}\,
\left(\cos\theta\,{\sigma^x_{\alpha\beta}}
\,+\,i\sin\theta\,{\sigma^y_{\alpha\beta}}\right)
  (i \partial_y) \biggr] \chi
\label{Sf}
\end{eqnarray}
$s=\pm$ and $\chi$ has a particle-hole index, acted on by
Pauli matrices $\vec{\tau}$; and a spin index,
acted on by Pauli matrices $\vec{\sigma}$.
\begin{eqnarray}
\chi_{a\alpha}(\vec{k}) = \left[ \begin{array}{c}
\chi_{11} \\ \chi_{21} \\ \chi_{12} \\ \chi_{22} 
\end{array} \right] = \left[ \begin{array}{c}
c_{\vec{ k}_F+\vec{k}\uparrow}^{\vphantom\dagger} \\ 
c_{-\vec{ k}_F-\vec{k}\downarrow}^{\dagger} \\ 
c_{\vec{ k}_F+\vec{k}\downarrow}^{\vphantom\dagger} \\ 
-c_{-\vec{ k}_F-\vec{k}\uparrow}^{\dagger} 
\end{array}\right].
\end{eqnarray}
In action (\ref{Sf}) we have included the electromagnetic field
$A^c_\mu$ and spin vector potential $A^{\sigma}_{\mu}$
which couple to the conserved electric and $S_z$
currents.

When $hc/4e$ vortex-$\pi$ disclination
composites are gapped, the $Z_2$ symmetry
$\varphi \rightarrow \varphi + \pi$, 
$\theta \rightarrow \theta + \pi$ plays no role,
and the other terms in (\ref{eqn:S_tot}) may be written in the form.
\begin{equation}
{S_c} = \frac{1}{2}{\rho_c}
\int {d^2}x\,d\tau\,\,\,
{\left({\partial_\mu}\varphi - A^c_{\mu} \right)^2} 
\,\,
\end{equation}
and 
\begin{eqnarray}
{S_\sigma} = \frac{1}{2}{\rho_\sigma}
\int {d^2}x\,d\tau\,\,\,
{\left({\partial_\mu}\theta - A^\sigma_{\mu} \right)^2}
\end{eqnarray}

The conserved electric and $S_z$
currents are given by:
\begin{eqnarray}
j^{c,\sigma}_{\mu} &=& \frac{\delta S_{\rm tot}}{\delta A^{c,\sigma}_{\mu}}
\end{eqnarray}
Conservation of charge and the $z$-component of spin
require
\begin{eqnarray}
\partial_{\mu} j^{c,\sigma}_{\mu} =0
\label{sconserv}
\end{eqnarray}

The interactions between the Goldstone fields and
the quasiparticles are highly non-linear in (\ref{eqn:Gold-qp-int1}).
This interaction can be made more tractable,
following \cite{Balents98}, if we define new fermion
fields $\psi$:
\begin{eqnarray}
\chi = {e^{i\varphi {\tau^z}/2}}\,
{e^{i\theta {\sigma^z}/2}}\, \psi
\label{psiintro}
\end{eqnarray}

With this change of variables, we have defined a
neutral, spinless fermion, $\psi$, which is governed by
the action:
\begin{eqnarray}
\label{eqn:Gold-qp-int2}
{S_f}  &=&  \int {d^2}x\,d\tau\,\biggl(\,
\psi^\dagger [ \partial_\tau - v_F \tau^z
  i\partial_x -
  v_{\scriptscriptstyle\Delta} {\tau^x}
{\sigma^x}\,  (i \partial_y) ] \psi\cr
& & {\hskip 1.0 cm}+\,\frac{1}{2}\,
\psi^\dagger [{\tau^z} {\partial_\tau}\varphi
- 2{A^c_\tau}{\tau^z}
- {v_F} {\partial_x}\varphi + 2{v_F} {A^c_x}] \psi\cr
& & {\hskip -0.5 cm}\,+\,\frac{1}{2}\, \psi^\dagger 
[{\sigma^z} {\partial_\tau}\theta
- 2{A^\sigma_\tau}{\tau^z}
- {v_F} {\tau^z}{\sigma^z}
{\partial_x}\theta + 2{v_F} {\tau^z}{\sigma^z}{A^\sigma_x}] \psi
\biggr)
\end{eqnarray}

The couplings between the Goldstone modes
and the quasiparticles are now either
trilinear or biquadratic
\begin{eqnarray}
\label{eqn:Gold-qp-int2a}
{S_f}  = {S^0_f} &+&
\frac{1}{2}\int {d^2}x\,d\tau\,\biggl(\,
J_0^c ( \partial_{\tau} \varphi - 2 A^c_{\tau} ) \cr
&+& J_x^c ( \partial_x \varphi - 2 A^c_{x} )
+
J_0^\sigma ( \partial_{\tau} \theta - 2 A^\sigma_{\tau} )
\cr
&+& J_x^\sigma ( \partial_x \theta - 2 A^\sigma_{x} )
\biggr)
\end{eqnarray}
with 
\begin{eqnarray}
\begin{array}{ll}
J_0^c = \psi^\dagger \tau^z \psi \hspace{0.5cm} &
J_x^c = - v_F \psi^\dagger  \psi \hspace{0.5cm} \\
J_0^{\sigma} = \psi^\dagger \sigma^z \psi \hspace{2cm} &
J_x^{\sigma} = - v_F \psi^\dagger  \sigma^z \tau^z\psi \hspace{2cm} \\
\end{array} \hspace{-2cm}
\label{currents}
\end{eqnarray}
The price that must be
paid is that the change of variables
(\ref{psiintro}) is not single-valued about
a topological defect. In particular, the charge part,
$\exp({i\varphi {\tau^z}/2})$, is double-valued
under transport about a flux $hc/2e$ vortex
since $\varphi$ winds by $2\pi$, while
the spin part, $\exp({i\theta {\sigma^z}/2})$, is double-valued
under transport about a meron
since $\theta$ winds by $2\pi$.

As we will see below, $\psi$s are weakly-coupled
quasiparticles in those quantum disordered
phases in which flux $hc/2e$ vortices and merons
are gapped.

\subsection{Defect Condensation}
\label{condensation}

Defect condensation is now implemented with
dual representations for the order parameters.
\cite{Balents98,Balents99,Balents00,dual}.
In the dual description of the $XY$ model \cite{dual},
the ordering field, $\varphi$,
is replaced by a gauge field which
parametrizes the total current,
together with a vortex field which accounts for
the singularities of $\varphi$.

We use the conservation of charge to define the dual gauge field
\begin{eqnarray}
{\epsilon_{\mu\nu\lambda}}{\partial_\nu}{a_\lambda^c}
= J^{tot\,\,c}_{\mu} = 
\rho_c \left({\partial_\mu}\varphi - A^c_{\mu} \right)
+ J^c_{\mu}
\end{eqnarray}
with $J^c_{\mu}$ from (\ref{currents}),
and introduce the vortex current
\begin{eqnarray}
j_{\mu}^{v} = \frac{1}{2\pi} {\epsilon_{\mu\nu\lambda}}
{\partial_\nu}{\partial_\lambda} \varphi
\label{jvdef}
\end{eqnarray}
which is not vanishing for a multivalued $\varphi$.
With the last two equations we can relate
the vortex current to the dual gauge field $a_{\mu}^c$
and quasiparticle current $J^c_{\mu}$
\begin{eqnarray}
j_{\mu}^{v} = {\epsilon_{\mu\nu\lambda}}
{\partial_\nu} \left[ \rho_c^{-1} 
{\epsilon_{\lambda\alpha\beta}} \partial_{\alpha} a^c_{\beta}
+ A^c_{\lambda} - \rho_c^{-1} J^c_{\lambda} \right]
\label{jv}
\end{eqnarray} 
Now a dual action for the charged degrees of
freedom is easily constructed by requiring
that its
equations of motion reproduce (\ref{jv}).
\begin{eqnarray}
S^c_{Dual} &=& S_{GL}({\Phi_v},a^c_{\mu})\, +\cr 
& & \int d\tau\,{d^2}x\, \left(  \frac{1}{2 \rho_c} (f^c_{\mu \nu})^2
+ a_{\mu}^c {\epsilon_{\mu\nu\lambda}}
{\partial_\nu} ( A^c_{\lambda} - {1 \over \rho_c} J^c_{\lambda}) \right) 
\end{eqnarray}
where 
\begin{eqnarray}
{S_{GL}}\left[\Phi,{a_\mu}\right] &=& \int d\tau\,{d^2}x\,\left(
{\rho_d \over 2} |(\partial_\mu - i a_\mu)\Phi |^2
+ V(\Phi)   \right)
\label{Scdual}
\end{eqnarray} 
and $f^c_{\mu\nu} = \partial_{\nu} a^c _{\mu}- \partial_{\mu} a^c _{\nu}$.
The field $\Phi_v^{\dagger}$ may be thought of as a vortex creation field.
The vortex current is given by
\begin{eqnarray}
j_{\mu}^{v} = {\rho_d \over 2} \left[ 
\Phi^{\dagger}_v ( {1 \over i} \partial_{\mu} - a^c_{\mu} )
{\Phi_v} + h.c. \right]
\end{eqnarray} 

An identical construction is now used for $\theta$
with $\rho_c$ replaced by $\rho_\sigma$ and
$J^c_{\lambda}$ by $J^\sigma_{\lambda}$:
\begin{eqnarray}
\label{Ssdual}
S^{\sigma}_{Dual} &=& S_{GL}({\Phi_m},a^{\sigma}_{\mu})\, +\cr 
& & \int d\tau\,{d^2}x\, \left( 
\frac{1}{2 \rho_\sigma} (f^\sigma_{\mu \nu})^2
+ a_{\mu}^\sigma {\epsilon_{\mu\nu\lambda}}
{\partial_\nu} ( A^\sigma_{\lambda} - {1 \over \rho_c} 
J^\sigma_{\lambda}) \right) 
\end{eqnarray}
where $\Phi^{\dagger}_m$ is a meron
creation operator. Analogous topological objects in the spin sector
have been discussed in the context of quantum Hall systems \cite{Lee,Moon,QH1}
and quantum antiferromagnets \cite{Read89,Sachdev94,Ng99a,Ng99b}.

Actions (\ref{Scdual}) and (\ref{Ssdual}) need to be
supplemented
by Chern-Simons gauge fields which enforce
the minus sign which is acquired when a $\psi$
encircles a flux $hc/2e$ vortex or a meron \cite{Balents98,Nayak00b}. 
With these additions, we obtain the following dual action
\begin{eqnarray}
S_{Dual} &=& S_{GL}(\Phi_v,a^c_{\mu} - a^{s1}_{\mu}) + 
S_{GL}(\Phi_m,a^\sigma_{\mu} - a^{s2}_{\mu}) 
\nonumber\\
&+&\int d\tau\,{d^2}x\, \biggl(  \frac{1}{2\rho_c} (f^c_{\mu \nu})^2 
+ a_{\mu}^c {\epsilon_{\mu\nu\lambda}}
{\partial_\nu} ( A^c_{\lambda} - {1\over \rho_c} J^c_{\lambda}) \cr
& & {\hskip 1.5 cm}+\, 2 \alpha^1_{\mu} {\epsilon_{\mu\nu\lambda}}
{\partial_\nu}
a_{\lambda}^{s1} + \alpha^1_{\mu} J_{\mu}^c
\biggr) 
\nonumber\\
&+&  
\int d\tau\,{d^2}x\, \biggl(  \frac{1}{2\rho_{\sigma}} (f^\sigma_{\mu \nu})^2 
+ a_{\mu}^\sigma {\epsilon_{\mu\nu\lambda}}
{\partial_\nu} ( A^{\sigma}_{\lambda} - {1 \over \rho_{\sigma}}
J^{\sigma}_{\lambda}) \cr
& & {\hskip 1.5 cm}+\,2 \alpha^2_{\mu} {\epsilon_{\mu\nu\lambda}}
{\partial_\nu} a_{\lambda}^{s2} + \alpha^2_{\mu} J_{\mu}^{\sigma}
\biggr) 
\end{eqnarray} 
where $\alpha_{\mu}^{1,2}$ and $a_\mu^{s1,2}$
are the gauge fields that perform 
the flux attachement and enforce the minus sign.

With this action in hand, we can now
address the quantum disordered phases
and quantum number separation. In essence,
there are {\it three} quantum numbers:
charge, spin, and electron number modulo 2.
These can separate in a variety
of patterns.

\begin{itemize}

\item If $\langle{\Phi_v}\rangle\neq 0$, flux $hc/2e$ vortices
condense. The Meissner
effect associated with this condensate imposes
\begin{equation}
{a^c_\mu} + {a^{s1}_\mu} = 0
\end{equation}
Recalling that $\epsilon_{\alpha\beta} \partial_{\alpha} a^c_{\beta} =
J_0^{tot~c}$ ($\alpha$, $\beta = x$,$y$)  is the charge density and
 $\epsilon_{\alpha\beta} \partial_{\alpha} a^{s1}_{\beta} =
J_0^{c}$  is the quasiparticle density,
we conclude that in this phase charge is attached to the $\psi$s.

\item If $\langle{\Phi_{m}}\rangle\neq 0$, merons condense,
and the Meissner effect associated with this condensate imposes
\begin{equation}
{a^{\sigma}_\mu} + {a^{s2}_\mu} = 0
\end{equation}
As $\epsilon_{\alpha\beta} \partial_{\alpha} a^\sigma_{\beta} =
J_0^{tot~\sigma}$ is the local spin density 
and $\epsilon_{\alpha\beta} \partial_{\alpha} a^{s2}_{\beta} =
J_0^{\sigma}$  we find that spin is attached to the $\psi$s.
All the fermions carry spin.

\item If $\langle{\Phi_{m}}\rangle = 0$, $\langle{\Phi_v}\rangle = 0$,
but $\langle{\Phi_v}{\Phi_{m}}\rangle\neq 0$
$hc/2e$ vortex - meron composites condense.
The Meissner effect associated with this
condensate imposes
\begin{equation}
{a^c_\mu} + {a^{\sigma}_\mu} = 0
\end{equation}
In other words, spin and charge are confined, but
the fermion $\psi$ carries neither since $\psi$ does not
acquire any phase upon encircling this composite
object, as evinced by the fact that ${\Phi_v}{\Phi_{m}}$
is not coupled to statistical gauge fields.

\item The condensation of other composites, such
as ${\Phi_{m}^2}$ (i.e. skyrmions), ${\Phi^2_v}$, etc.
does not cause the confinement
of any quantum numbers.

\end{itemize}

\subsection{Exotic Phases}
\label{EN}

The order parameter classification
discussed after equation (\ref{Qop})
is incomplete; those states can
occur in several varieties, classified by
the allowed quantum numbers \cite{endnote1}.

{\it Charge-$4e$ singlet superconductors},
${\Delta^{4e}}\neq 0$, $Q = 0$.
\begin{trivlist}
\item { 1A)}
If $\langle {\Phi_{m}}\rangle \neq 0$,
then the fermionic excitations carry spin $1/2$.
\item{ 1B)} However, if $\langle {\Phi_{m}}\rangle = 0$
but $\langle (\Phi_m)^2 \rangle \neq 0$
then the $\psi$s are spinless.
Note that the charge quantum number of the fermionic excitation
is not really well-defined since $U(1)$ is broken
in the superconducting state; said differently,
the fermion can always exchange charge with the
condensate.
\end{trivlist}

{\it  Spin-triplet insulator}, ${\Delta^{4e}}= 0$,
$Q \neq 0$. 
\begin{trivlist}
\item { 2A)} If $\langle {\Phi_{v}}\rangle \neq 0$,
the $\psi$s carry charge $e$.
\item { 2B)} If $\langle {\Phi_{v}}\rangle = 0$ but
$\langle {\left({\Phi_{v}}\right)^2}\rangle \neq 0$,
then the $\psi$s are neutral. As in the previous case,
the spin quantum number of the $\psi$s is not well-defined.
\end{trivlist}

{\it Spin-singlet insulator}, 
${\Delta^{4e}}= 0$,
$Q = 0$. 
\begin{trivlist}
\item 
{ 3A) } If $\langle {\Phi_{v}}\rangle \neq 0$,
$\langle {\Phi_{m}}\rangle \neq 0$,
then the $\psi$s carry spin $1/2$ and charge $e$.
Phase CSf.
\item
{3B) } If $\langle {\Phi_{m}}\rangle = 0$
but $\langle ({\Phi_{m}})^2\rangle \neq 0$
while $\langle {\Phi_{v}}\rangle \neq 0$, then
the $\psi$s are charge $e$, spinless fermionic
excitations: {\it CfSb} Phase.
\item
{3C) }  If $\langle {\Phi_{v}}\rangle = 0$ but
$\langle {\left({\Phi_{v}}\right)^2}\rangle \neq 0$
while $\langle {\Phi_{m}}\rangle \neq 0$,
then the $\psi$s are neutral, spin $1/2$ fermionic
excitations: {\it CbSf} Phase.
\item
{3D) } If  $\langle {\Phi_{v}}{\Phi_{m}}\rangle \neq 0$,
then the $\psi$s are neutral, spinless fermionic
excitations, but spin and charge are confined into
a bosonic spin $1/2$, charge $e$ excitation:
{\it CSbNf} Phase.
\item
{3E) } Finally, if $\langle {\Phi_{v}}\rangle = 0$ but
$\langle {\left({\Phi_{v}}\right)^2}\rangle \neq 0$
and $\langle {\Phi_{m}}\rangle = 0$
but $\langle ({\Phi_{m}})^2\rangle \neq 0$,
then the $\psi$s are neutral, spinless fermionic
excitations. Bosonic charge $e$ excitations, ${e^{i\varphi/2}}$,
and bosonic spin $1/2$ excitations, ${e^{i\theta/2}}$
are also liberated: {\it CbSbNf} Phase.
\end{trivlist}

To summarize, we have the following phases
with exotic quantum numbers.
\begin{itemize}

\item A charge-$4e$ singlet superconductor
with spinless fermionic excitations.

\item A spin-triplet insulator with
neutral fermionic excitations.

\item Spin-singlet insulators with (a) charge $e$
spinless fermions and spin $1/2$ neutral 
bosons; 
(b) spin $1/2$ neutral
fermions and spinless charge $e$ bosons; 
(c) neutral spinless fermions, bosonic charge $e$ 
spinless excitations, and bosonic spin $1/2$ neutral
excitations; or (d) neutral spinless fermionic
excitations and bosonic charge $e$ spin $1/2$
excitations.

\end{itemize}

These result are summarized in the following
diagrams that describe various phases that can result
from quantum disordering a $p$-wave superconductor.


Scenario proposed in this section for quantum number separation in
$p$-wave superconductors may apply to other systems, provided that
they acquire non-trivial topological order in the spin and charge
sectors, or in the language of this section when they have
sufficiently strong quantum fluctuations of spin and charge degrees of
freedom simultaneously.  In Appendix \ref{dSCAF} we show that quantum
disordered $d$-wave superconductor with easy-plane antiferromagnetic
fluctuations may be treated in the same way as we treated $p$-wave
superconductors in this section.

\section{$Z_2 \times Z_2$ Lattice Gauge Theory. }

In this section, we derive a ${Z_2}\times{Z_2}$
gauge theory representation of a model
which gives rise to local $p$-wave superconducting
fluctuations. In addition to the superconducting
state, we find the exotic phases discussed in the
previous section. These have a simple description
as the various deconfining phases of the ${Z_2}\times{Z_2}$
gauge theory. Readers who are uninterested in
the technical details of our derivation may skip
directly to equations (\ref{S}), (\ref{S_terms}),
and the subsequent discussion.

\subsection{General Formalism.}

We consider the following Hamiltonian that 
describes the equal spin pairing state of a 
$p$-wave superconductor.
\begin{equation}
H = H_t + H_u + H_v + H_{\Delta} 
\end{equation}
with
\begin{eqnarray}
H_t &=& - t \sum_{r r', \alpha} 
c^{\dagger}_{r \alpha} c_{r' \alpha} + h.c. \ , \cr
H_u &=& u \sum_r (N_r - N_0)^2 \ , \cr 
H_v &=& v \sum_r (M_r)^2 \ , \cr
H_{\Delta} &=& 
\sum_{r r'} [\Delta^{\uparrow \uparrow}_{r r'} 
c_{r \uparrow} c_{r' \uparrow} + 
\Delta^{\downarrow \downarrow}_{r r'} 
c_{r \downarrow} c_{r' \downarrow}]
+ h.c. \ , 
\end{eqnarray}
where $\Delta^{\uparrow \uparrow}_{r r'}$ and 
$\Delta^{\downarrow \downarrow}_{r r'}$ represent
the order parameter fields for the Cooper pairs with 
spin up-up and down-down pairs respectively.
Here $\alpha = \uparrow, \downarrow$ is the spin index.
The term proportional to $u$ represents the on-site
Coulomb repulsion. $N_r$ is the total number operator 
of electrons at the site $r$, $N_0$ is the average 
electron number per site. $M_r$ is the $z$-component 
of the total spin operator.
At equilibrium, $|\Delta^{\uparrow \uparrow}|=
|\Delta^{\downarrow \downarrow}|$.
Note that there are two independent phases associated
with $\Delta^{\uparrow \uparrow}$ and
$\Delta^{\downarrow \downarrow}$.
We can rewrite $H_{\Delta}$ as
\begin{equation}
H_{\Delta} = \Delta \sum_{r r'} a_{r r'} [ 
e^{i \varphi_{r \uparrow}} c_{r \uparrow} 
c_{r' \uparrow} 
+ e^{i \varphi_{r \downarrow}} 
c_{r \downarrow} c_{r' \downarrow}]
+ h.c. \ , 
\end{equation}
where $\Delta = |\Delta^{\uparrow \uparrow}|
=|\Delta^{\downarrow \downarrow}|$ and $a_{r r'}$
is the form factor that gives rise to the particular 
$p$-wave symmetry. 

The fields $\varphi_{r \uparrow}$ and
$\varphi_{r \downarrow}$ are canonically conjugate to
the Cooper pair number operators of up-up and 
down-down Cooper pairs, $n_{r \uparrow}$
and $n_{r \downarrow}$:
\begin{equation}
[\varphi_{r \uparrow}, n_{r' \uparrow}] 
= i \delta_{r r'} \ , \ \ \ 
[\varphi_{r \downarrow}, n_{r' \downarrow}] 
= i \delta_{r r'}  \ . 
\end{equation}
The conserved charge densities for the electrons
with spin $\uparrow$ and $\downarrow$ are given by
\begin{eqnarray}
N_{r \uparrow} &=& 2 n_{r \uparrow} + 
\rho_{r \uparrow} \cr
N_{r \downarrow} &=& 2 n_{r \downarrow} + 
\rho_{r \downarrow} \ ,
\end{eqnarray}
where $\rho_{r \alpha} = 
c^{\dagger}_{r \alpha} c_{r \alpha}$
is the quasiparticle number, which is not
equal to the electron number. It is
useful to remind the readers that 
the Hamiltonian (64) does not conserve
the quasiparticle number, since it contains 
terms that annihilate a pair of them
and create a Cooper pair.
Only the total number of electrons 
of a given spin, given by equation (67),
is conserved. This may  be 
formulated as a conservation of
the total charge and $z$-component of
the total spin.
\begin{equation}
N_{r} = N_{r \uparrow} + N_{r \downarrow}, \ \ \
M_{r} = N_{r \uparrow} - N_{r \downarrow} \ .
\end{equation}

Let us define boson operators, $b_{r \alpha}$, which
carry charge $e$ and spin $\alpha = \uparrow, \downarrow$:
\begin{equation}
b^{\dagger}_{r \alpha} = 
t^{\alpha}_r e^{i \varphi_{r \alpha} / 2}
= e^{i \phi_{r \alpha}} \ , 
\end{equation}
where $t^{\alpha}_r = \pm 1$ are Ising variables
and $\phi_{r \alpha}$ are defined in the interval
zero to $2 \pi$. 
Note that the squares of $b^{\dagger}_{r \uparrow}$ and 
$b^{\dagger}_{r \downarrow}$ create
the spin up-up and down-down Cooper pairs via 
the following relation.
\begin{equation}
(b^{\dagger}_{r \alpha})^2 = e^{i \varphi_{r \alpha}} \ .
\end{equation} 
One can also see that the canonical conjugates 
of $\phi_{r \alpha}$ are the total densities of
electrons with spin $\uparrow$ and $\downarrow$.
They satisfy the following commutation relations.
\begin{equation}
[\phi_{r \alpha}, N_{r' \alpha}] 
= i \delta_{r r'} \ .
\end{equation}
Similarly, the following commutation relations are 
also satisfied.
\begin{equation}
[\phi_{r c}, N_{r'}] 
= i \delta_{r r'}, \ \ \ 
[\phi_{r s}, M_{r'}] 
= i \delta_{r r'} \ ,  
\end{equation}
where $\phi_{r c} = (\phi_{r \uparrow}+\phi_{r \downarrow})/2$
and $\phi_{r s} = (\phi_{r \uparrow}-\phi_{r \downarrow})/2$.
At this stage, it is useful to define the 
fermion operators, $f^{\dagger}_{r \alpha}$, as follows.
\begin{equation}
c^{\dagger}_{r \alpha} = b^{\dagger}_{r \alpha} 
f^{\dagger}_{r \alpha} \ .
\end{equation}
Note that $f^{\dagger}_{r \alpha}$ creates spinless
neutral fermions due to the fact that 
$b^{\dagger}_{r \alpha}$ carries both the charge
and spin of the electrons.

It is also useful to define $\varphi_{r c}$ and 
$\varphi_{r s}$
as follows.
\begin{equation}
e^{i \varphi_{r \uparrow}} = 
e^{i \varphi_{r c}} e^{i \varphi_{r s}},  \ \ 
e^{i \varphi_{r \downarrow}} = 
e^{i \varphi_{r c}} e^{-i \varphi_{r s}} \ . 
\end{equation} 
Note that there is a $Z_2$ symmetry associated with
these definitions of phase variables; 
$\varphi_{r c} \rightarrow \varphi_{r c} + \pi$
and $\varphi_{r s} \rightarrow \varphi_{r s} + \pi$
do not change $e^{i \varphi_{r \uparrow}}$
and $e^{i \varphi_{r \downarrow}}$. 
Now we define boson operators, $b^{\dagger}_r$ and 
$z^{\dagger}_r$, as
\begin{equation}
b^{\dagger}_r = t_r e^{i \varphi_{r c}/2}
= e^{i \phi_{r c}} \ ,  \ \ 
z^{\dagger}_{r} = s_r e^{i \varphi_{r s}/2}
= e^{i \phi_{r s}} \ . 
\end{equation}
Here $t_r = \pm 1$ and $s_r = \pm 1$
are Ising variables.
Note that these operators satisfy the following 
identities.
\begin{equation}
(b^{\dagger}_r)^2 = e^{i \varphi_{r c}},  \ \ 
(z^{\dagger}_r)^2 = e^{i \varphi_{r s}} \ .  
\end{equation}
Note also that $b^{\dagger}_{r \uparrow}$ and
$b^{\dagger}_{r \downarrow}$ can be rewritten as
\begin{equation}
b^{\dagger}_{r \uparrow} = b^{\dagger}_r z^{\dagger}_r,  \ \
b^{\dagger}_{r \downarrow} = b^{\dagger}_r z_r \ .
\end{equation}

Now the total Hamiltonian can be written as
\begin{equation}
H = H_t + H_u + H_v + H_{\Delta} 
\end{equation}
with
\begin{eqnarray}
H_t &=& - t \sum_{r r'} ( 
b^{\dagger}_r b_{r'} z^{\dagger}_r z_{r'}
f^{\dagger}_{r \uparrow} f_{r' \uparrow} + 
b^{\dagger}_r b_{r'} z_r z^{\dagger}_{r'}
f^{\dagger}_{r \downarrow} f_{r' \downarrow} ) \cr
& & \,+ \,\,h.c.\cr
H_{\Delta} &=& 
\Delta \sum_{r r'} a_{r r'} (  
b^{\dagger}_r b_{r'} z^{\dagger}_r z_{r'}
f_{r \uparrow} f_{r' \uparrow}
+ b^{\dagger}_r b_{r'} z_r z^{\dagger}_{r'}
f_{r \downarrow} f_{r' \downarrow}) \cr
& & \,+\, h.c.
\label{refeq}
\end{eqnarray}
The Hamiltonian is invariant under
the following local
transformations:
\begin{trivlist}
\item $ {\rm i)}$ $ Z_{2\tau}$: $ b_r \rightarrow -b_r$; 
$f_{r \alpha} \rightarrow -f_{r \alpha}$
\item ${\rm ii)}$  $Z_{2\sigma}$:  $z_r \rightarrow -z_r$; 
$f_{r \alpha} \rightarrow -f_{r \alpha}$;
\item ${\rm iii)}$  $Z_{2\tilde{\sigma}}$:  $z_r \rightarrow -z_r$; 
$b_{r \alpha} \rightarrow -b_{r \alpha}$
\end{trivlist}
Only two of these transformations are independent,
any one of them can be represented as a product of the other 
two. Together they form  $Z_2 \times Z_2$ gauge symmetry,
that has three $Z_2$ subgroups as reflected 
in three possible transformations above.
These subgroups are distinct, 
but not independent.
$Z_2 \times Z_2$ local gauge symmetry is a consequence of the
redundancy in the enlarged Hilbert space of
$f_{r \alpha}$, $b_r$, and $z_r$.
There is a further redundancy in our
description in terms of $b_r$ and
$z_r$  because $b_r \rightarrow i\, b_r$, $z_r \rightarrow -i\, z_r$
also leaves all physical quantities invariant.
This identification allows for the existence
of flux $\pi$-disclination-$hc/4e$ vortex composites
which we discussed in subsection \ref{td}.
As before, we assume that these topological defects
are gapped so that we can safely ignore this
identification and take $\phi_{rc}$ and $\phi_{rs}$
as defined from $[0,2 \pi)$.

In order to get the correct Hilbert space of the electrons,
we have to impose two constraints at each site.
\begin{eqnarray}
N_r + \rho_{r\uparrow} + \rho_{r \downarrow} &=& 
{\rm even \ number},  \cr
M_r + \rho_{r\uparrow} - \rho_{r \downarrow} &=& 
{\rm even \ number} \ .
\end{eqnarray}
These can be written as
\begin{equation}
(-1)^{N_r + \rho_{r\uparrow} + \rho_{r \downarrow}} = 1,
\ \ (-1)^{M_r + \rho_{r\uparrow} - 
\rho_{r \downarrow}} = 1 \ .
\end{equation}  
The constraints can be implemented in the path integral
representation of the partition function using the
following projection operators.
\begin{equation}
{\cal P}_{c} = \prod_r 
{\cal P}_{r c}, \ \ 
{\cal P}_{s} = 
\prod_r {\cal P}_{r s}   
\end{equation}
with 
\begin{eqnarray}
{\cal P}_{r c} &=& {1 \over 2}[1 +
(-1)^{N_r + \rho_{r\uparrow} + \rho_{r \downarrow}}]\cr
 &=& {1 \over 2} \sum_{\sigma_r=\pm 1}
e^{i {\pi \over 2}(1-\tau_r)
(N_r + \rho_{r\uparrow} + \rho_{r \downarrow})} \cr
{\cal P}_{r s} &=& {1 \over 2}[1 +
(-1)^{M_r + \rho_{r\uparrow} - 
\rho_{r \downarrow}}] \cr
&=& 
{1 \over 2} \sum_{\tau_r=\pm 1}
e^{i {\pi \over 2}(1-\sigma_r)
(M_r + \rho_{r\uparrow} - 
\rho_{r \downarrow})} \ .
\end{eqnarray}

Using the projection operators, the partition function 
can be written as
\begin{equation}
Z = Tr[e^{-\beta H} {\cal P}_c
{\cal P}_s] \ .
\end{equation}
A Euclidean path integral representation can
be obtained by splitting the exponential
into $M$ number of time slices.
\begin{equation}
Z = Tr[(e^{-\epsilon H} {\cal P}_c
{\cal P}_s)^M] \ ,
\end{equation}
where $\epsilon = \beta/M$. 
Now the partition function can be written as
\begin{equation}
Z = \int \prod_{i \alpha} 
d{\bar f}_{i \alpha} df_{i \alpha}
d\phi_{i c} d\phi_{i s}
\sum_{N_i = -\infty}^{\infty}
\sum_{M_i = -\infty}^{\infty} 
\sum_{\sigma_i = \pm 1}
\sum_{\tau_i = \pm 1} e^{-S} \ .
\end{equation}
Here $i = (r,\tau)$ runs over the 2+1 dimensional
space-time lattice with $\tau = 1, 2, ... , M$ time
slices. The action, $S$, has the following form
\begin{equation}
S = S^f_{\tau} + S^{\phi_c}_{\tau} + S^{\phi_s}_{\tau}
+ \epsilon 
\sum_{\tau=1}^{M} H (N_{\tau}, M_{\tau}, \phi_{\tau c},
\phi_{\tau s}, {\bar f}_{\tau \alpha},f_{\tau \alpha}) \ 
\end{equation}
with
\begin{eqnarray}
S^f_{\tau} &=& \sum_{r,\tau=1}^{M} \sum_{\alpha} [ 
{\bar f}_{\tau \alpha} (\sigma_{\tau + 1} \tau_{\tau + 1}
f_{\tau+1,\alpha} - f_{\tau \alpha}) ] \ , \cr
S^{\phi_c}_{\tau} &=& \sum_{r,\tau=1}^{M} 
N_{\tau} [\phi_{\tau c} - \phi_{\tau - 1,c}
+ {\pi \over 2} (1 - \tau_{\tau})] \ , \cr
S^{\phi_s}_{\tau} &=& \sum_{r,\tau=1}^{M}
M_{\tau} [\phi_{\tau s} - \phi_{\tau - 1,s}
+ {\pi \over 2} (1 - \sigma_{\tau})] \ , \cr
\end{eqnarray}
Here the spatial index $r$ is suppressed for clarity.
The Ising variables $\sigma_{\tau}$ and
$\tau_{\tau}$ are defined on the links 
connecting adjacent time slices and can be regarded as
the time component of the Ising gauge fields.

The sum of $H_t$ and $H_{\Delta}$, can be decoupled 
using the Hubbard-Stratanovich
fields $\chi_{r r'}$ and $\eta_{r r'}$.
\begin{equation}
e^{-\epsilon (H_t + H_{\Delta})} = \int \prod_{r r'} 
\prod_{\tau} 
d\chi_{rr'} d\chi^*_{rr'} 
d\eta_{rr'} d\eta^*_{rr'}
e^{-S_{t,\Delta}} \ . 
\end{equation}
Using the expressions for $H_t$ and $H_{\Delta}$
\begin{eqnarray}
H_t &=& - t \sum_{r r', \alpha} ( 
b^{\dagger}_{r \alpha} b_{r' \alpha} 
f^{\dagger}_{r \alpha} f_{r' \alpha} + 
h.c.) \ , \cr
H_{\Delta} &=& 
\Delta \sum_{r r', \alpha} a_{r r'} (  
b^{\dagger}_{r \alpha} b_{r' \alpha} 
f_{r \alpha} f_{r' \alpha} + 
h.c.) \ .
\end{eqnarray}
we have
\begin{eqnarray}
S_{t,\Delta} &=& {1 \over 4} 
\epsilon \sum_{rr', \alpha} \biggl\{ 
\Bigl[ 2 |\chi_{rr'}|^2 - \chi_{rr'} 
(b^*_{r \alpha} b_{r' \alpha} +
t {\bar f}_{r \alpha} f_{r' \alpha}\cr
& & {\hskip 1.6 cm}+ a_{rr'} \Delta f_{r \alpha} f_{r' \alpha})\Bigr] \cr
& & {\hskip 1.0 cm}+ \Bigl[2 |\eta_{rr'}|^2 - \eta_{rr'} 
(b^*_{r \alpha} b_{r' \alpha} -
t {\bar f}_{r \alpha} f_{r' \alpha}\cr
& & {\hskip 1.8 cm} - a_{rr'} \Delta f_{r \alpha} f_{r' \alpha})\Bigr]
+ c.c. 
\biggr\} \ .
\end{eqnarray}
Rearranging terms, we get
\begin{eqnarray}
S_{t,\Delta} &=& {1 \over 4} \epsilon \sum_{rr', \alpha} 
\biggl[ 2 |\chi_{rr'}|^2
+ 2 |\eta_{rr'}|^2  \cr
& & {\hskip 1.2 cm}
- (\chi_{rr'} + \eta_{rr'}) 
b^*_{r \alpha} b_{r' \alpha}\cr
& & {\hskip 1.2 cm}
- t (\chi_{rr'} - \eta_{rr'}) 
{\bar f}_{r \alpha} f_{r' \alpha} \cr
& &{\hskip 1.2 cm}
- a_{rr'} \Delta (\chi_{rr'} - \eta_{rr'}) 
f_{r \alpha} f_{r' \alpha}
+ c.c. \biggr] \ .
\end{eqnarray}
Rewriting this in terms of $b_r$ and $z_r$, we get
\begin{eqnarray}
S_{t,\Delta} &=& {1 \over 4} \epsilon \sum_{rr'} \biggl\{ 
\sum_{\alpha} \Bigl[ 2 |\chi_{rr'}|^2 + 2 |\eta_{rr'}|^2 \cr
& & {\hskip 1.6 cm}
- t(\chi_{rr'} - \eta_{rr'}) {\bar f}_{r \alpha} 
f_{r' \alpha}\cr
& & {\hskip 1.6 cm}
- a_{rr'} \Delta (\chi_{rr'} - \eta_{rr'}) 
f_{r \alpha} f_{r' \alpha} \Bigr] \cr 
& & - (\chi_{rr'} + \eta_{rr'}) (b^{\dagger}_r b_{r'} z^{\dagger}_r
z_{r'} + b^{\dagger}_r b_{r'} z_r z^{\dagger}_{r'} ) 
+ c.c. \biggr\} \ .
\end{eqnarray}
In order to decouple $b_r$ from $z_r$, another 
Hubbard-Stratanovich transformation is necessary.
Using similar procedures, the following term
\begin{equation}
- {1 \over 4} \epsilon \sum_{rr'} [
(\chi_{rr'} + \eta_{rr'}) 
(b^{\dagger}_r b_{r'} z^{\dagger}_r
z_{r'} + b^{\dagger}_r b_{r'} z_r z^{\dagger}_{r'}) 
+ c.c. ] \ .
\end{equation}
can be decoupled as
\begin{eqnarray}
- {1 \over 16} \epsilon \sum_{rr'} 
(\chi_{rr'} + \eta_{rr'}) 
\biggl[ 2 |\lambda_{rr'}|^2
+ 2 |\xi_{rr'}|^2 +
2 |p_{rr'}|^2\cr
+ 2 |q_{rr'}|^2 
- (\lambda_{rr'} + \xi_{rr'}) 
z^{\dagger}_r z_{r'}
- (p_{rr'} + q_{rr'}) 
z_r z^{\dagger}_{r'}\cr
- (p_{rr'} - q_{rr'} + \lambda_{rr} - \xi_{rr'}) 
b^{\dagger}_r b_{r'}
+ c.c. \biggr] \ ,
\end{eqnarray}
We now make a saddle point approximation and keep the
Ising fluctuations around this saddle point.
The natural choices are
\begin{eqnarray}
\chi_{rr'} - \eta_{rr'} 
&=& \sigma_{rr'} \tau_{rr'} \chi_f \ , \cr 
(\chi_{rr'} + \eta_{rr'}) 
(\lambda_{rr'} - \xi_{rr'} + p_{rr'} - q_{rr'}) 
&=&
\tau_{rr'} \chi_c \ , \cr
(\chi_{rr'} + \eta_{rr'})
(\lambda_{rr'} + \xi_{rr'} + p^*_{rr'} + q^*_{rr'})
&=&
\sigma_{rr'} \chi_s \ .
\end{eqnarray}          
where $\sigma_{rr'} = \pm 1$ and 
$\tau_{rr'} = \pm 1$
are Ising fluctuations.
We drop all of the constant terms and define the
following variables
\begin{equation}
t_f = {1 \over 4} t \chi_f,  \ \ 
t_c = {1 \over 16} \chi_c, \ \ 
t_s = {1 \over 16} \chi_s, \ \ 
t_\Delta = {1 \over 4} \Delta \chi_f 
\end{equation} 
to obtain
\begin{eqnarray}
S^{\rm eff}_{t,\Delta} &=& - \epsilon \sum_{rr'} 
\sum_{\alpha}\biggl[ \sigma_{rr'} \tau_{rr'}
( t_f {\bar f}_{r \alpha} f_{r' \alpha}
+ t_{\Delta} a_{rr'} f_{r \alpha} f_{r' \alpha} )\cr
& & {\hskip 1.6 cm}
+ t_c \tau_{rr'} b^{\dagger}_r b_{r'} +
t_s \sigma_{rr'}
z^{\dagger}_r z_{r'} + c.c.\biggr] \ .  
\end{eqnarray}

Combining all the results, the approximate full partition 
function can be written as
\begin{eqnarray}
{\tilde Z} &=& \int \prod_{i \alpha} 
d{\bar f}_{i\alpha} df_{i\alpha} d\phi_{ic} d\phi_{is}\,\,\times\cr
& & {\hskip 1.6 cm}
\sum^{\infty}_{N_i = -\infty}
\sum^{\infty}_{M_i = -\infty} 
\prod_{\langle i j \rangle}
\sum_{\sigma_{ij} = \pm 1}
\sum_{\tau_{ij} = \pm 1} 
e^{-S} \ ,
\end{eqnarray}
where $\sigma_{ij}$ and 
$\tau_{ij}$ are $Z_2$ gauge
fields living on the nearest neighbor links of the
space-time lattice. 
The total action, $S$, is given by
\begin{equation}
S = S^{f}_{\tau} + S^{\phi_c}_{\tau} +
S^{\phi_s}_{\tau} + 
S_{\Delta} + S_0 + S_u + S_v
\end{equation}
with
\begin{eqnarray}
S^{f}_{\tau} &=& -i \sum_{i,j=i+{\hat \tau}}
\sum_{\alpha} 
[{\bar f}_{i\alpha} (\sigma_{ij} \tau_{ij} f_{j\alpha} 
- f_{i\alpha})] \ , \cr
S^{\phi_c}_{\tau} &=& -i \sum_{i,j=i-{\hat \tau}}  
N_{i} [\phi_{ic} - \phi_{jc} + 
{\pi \over 2} (1-\tau_{ij})] \ , \cr
S^{\phi_s}_{\tau} &=& -i \sum_{i,j=i-{\hat \tau}}  
M_{i} [\phi_{is} - \phi_{js} + 
{\pi \over 2} (1-\sigma_{ij})] \ , \cr
S_{\Delta} &=& \epsilon \sum_{i,j=i+{\hat x}} 
t_{\Delta} \sigma_{ij} \tau_{ij}  
(a_{ij} f_{i \alpha} f_{j \alpha}
+ c.c. ) \ , \cr
S_0 &=& - \epsilon \sum_{i,j=i+{\hat x}} 
\sum_{\alpha} \Bigl[ t_f \sigma_{ij} \tau_{ij}
{\bar f}_{i \alpha} f_{j \alpha}
+ t_c \tau_{ij} b^*_i b_j\cr
& & {\hskip 2.2 cm} +
t_s \sigma_{ij}
z^*_i z_j + c.c.\Bigr] \ , \cr
S_u &=& \epsilon u \sum_i (N_i - N_0)^2 \ , \cr
S_v &=& \epsilon v \sum_i (M_i)^2 \ ,  
\end{eqnarray}  
where ${\hat \tau}$ and ${\hat x}$ represent the
time and spatial links. $a_{ij}=a_{rr'}$ on the spatial
links and zero otherwise.

Using the Poisson resummation formula, one can show that 
\begin{eqnarray}
\sum_{N_i} e^{-(S_u + S^{\phi_c}_{\tau})}
&=& e^{\sum_{i,j=i-{\hat \tau}} {1 \over 2\epsilon u} 
\tau_{ij} {\rm cos}(\phi_{ic}-\phi_{jc}) - S^{\sigma}_B} \ , \cr
\sum_{M_i} e^{-(S_v + S^{\phi_s}_{\tau})} \ 
&=& e^{\sum_{i,j=i-{\hat \tau}} {1 \over 2\epsilon v} 
\sigma_{ij} {\rm cos}(\phi_{is}-\phi_{js})} 
\end{eqnarray}
with
\begin{eqnarray}
S^{\tau}_B &=& -i N_0 \sum_{i,j=i-{\hat \tau}} 
[2 \pi l^{\tau}_{ij} - {\pi \over 2}(1-\tau_{ij})] \ .
\label{berry}
\end{eqnarray}
Here $l^{\tau}_{ij}$ is defined as
\begin{equation}
l^{\tau}_{ij} = Int \left [ {\Phi^c_{ij} \over 2 \pi} + 
{1 \over 2} \right ] 
\end{equation}
with $\Phi^c_{ij} = \phi_{ic} - \phi_{jc} + 
{\pi \over 2} (1-\tau_{ij})$ is the gauge invariant
phase difference across the temporal link.
$Int$ denotes the integer part. 
One can see that the Berry phase term for 
$\sigma_{ij}$ is absent. This is due to the fact that
we have equal amplitudes for up-up and down-down
pairing in the equal spin pairing state, analogous
to particle-hole symmetry in the charge sector.

Gathering these terms, the final form of the
action is given by
\begin{equation}
S = S_f + S_c + S_s + S^{\sigma}_B + S_g
\label{S}
\end{equation}
with
\begin{eqnarray}
\label{S_terms}
S_f &=& - \sum_{ij, \alpha} \sigma_{ij} \tau_{ij}
[ t^f_{ij} {\bar f}_{i \alpha} f_{j \alpha} 
+ {\tilde t}^{\Delta} a_{ij} 
f_{i\alpha} f_{j\alpha} + c.c. ]\cr
& & -  \sum_{i \alpha} {\bar f}_{i\alpha} f_{i\alpha} \ , \cr
S_c &=& - \sum_{ij} t^c_{ij} \tau_{ij} 
(b^*_i b_j + c.c.) \ , \cr
S_s &=& - \sum_{ij} t^s_{ij} \sigma_{ij} 
(z_i z_j + c.c. ) 
\ .
\end{eqnarray}
Here $t^c_{ij}$ is $\epsilon t_c$ on the spatial link 
and ${1 \over 4 \epsilon u}$ on the temporal
link. Similarly $t^s_{ij}$ is $\epsilon t_s$ on the spatial link 
and ${1 \over 4 \epsilon v}$ on the temporal
link. $t^f_{ij} = \epsilon t_f$ on the spatial
link and $t^f_{ij} = -1$ on the temporal link.
And ${\tilde t}^{\Delta} = \epsilon t_{\Delta}$.
The last term of (\ref{S}) corresponds to the Maxwell
terms for the $Z_2$ gauge fields, that we assume are
generated after we integrate out
excitations at high energies. 
\begin{eqnarray}
S_g = &-&K_1 \sum_\Box 
 \prod_\Box \sigma_{ij}
      -K_2 \sum_\Box \prod_\Box \tau_{ij} \nonumber\\
      &-&K_3 \sum_{\Box} \prod_{\Box} \sigma_{ij} \tau_{ij}
\label{SG}
\end{eqnarray}
These are the simplest terms
providing dynamics of the gauge fields that are consistent with the
gauge symmetries
\begin{eqnarray}
Z_{2\tau}: \, b_i \rightarrow t_i b_i;\,\,\, 
f_{i \alpha} \rightarrow t_i f_{i \alpha};\,\,\, 
\tau_{ij} \rightarrow t_i t_j \tau_{ij}
\nonumber\\
Z_{2\sigma}: \, z_i \rightarrow s_i z_i;\,\,\, 
f_{i \alpha} \rightarrow s_i f_{i \alpha};\,\,\, 
\sigma_{ij} \rightarrow s_i s_j \sigma_{ij}
\label{Z2transf}
\end{eqnarray}
where $t_i$ and $s_i$ are $\pm 1$.
In the future we will call any particle that transforms under
the first and the second transformations of (\ref{Z2transf}) as having
$Z_{2\tau}$ and $Z_{2\sigma}$ charges respectively.

\subsection{Spin Singlet Insulating Phases}

Before discussing possible spin singlet insulating phases of 
the combined action
(\ref{S}) - (\ref{SG}) it is useful
to review properties of a pure $Z_2 \times Z_2$ gauge theory (\ref{SG}).
Under duality transformation defined in \cite{Senthil99,Fradkin79,Wegner71}
this model becomes a generalized Ashkin-Teller model \cite{Domb77} 
\begin{eqnarray}
S_{AT} = &-&K_{d1} \sum_{\langle ij \rangle} v_i v_j
-K_{d2} \sum_{\langle ij \rangle} u_i u_j \nonumber\\
&-&K_{d3} \sum_{\langle ij \rangle} u_i v_i u_j v_j
\label{SAT}
\end{eqnarray}
Here $u_i$ and $v_i$ are Ising variables defined on the dual lattice
in $d=2+1$.
We can identify five possible phases of (\ref{SAT}):
\begin{trivlist}
\item 1) Fully ordered phase $\langle u \rangle \neq 0$,  
$\langle v \rangle \neq 0$,
 $\langle u v \rangle \neq 0$;
\item 2) Partially ordered phase $\langle u \rangle \neq 0$,  
$\langle v \rangle = 0$,
 $\langle u v \rangle = 0$;
\item 3) Partially ordered phase $\langle u \rangle = 0$,  
$\langle v \rangle \neq 0$,
 $\langle u v \rangle = 0$;
\item 4) Partially ordered phase $\langle u \rangle = 0$,  
$\langle v \rangle = 0$,
 $\langle u v \rangle \neq 0$;
\item 5) Disordered phase $\langle u \rangle = 0$,  
$\langle v \rangle = 0$, $\langle u v \rangle = 0$.
\end{trivlist}

As pointed out in \cite{Senthil99} the Ising variables of (\ref{SAT})
correspond to the $Z_2$ vortices of the original gauge model They
describe gauge field configurations with plaquette products equal to
minus one, i.e. plaquettes pierced by $Z_2$ fluxes. Following
\cite{Senthil99} we call such $Z_2$ vortices "visons". In fact we have
three kinds of visons: $\sigma$-visons that describe $Z_2$ vortices of
$\sigma$, $\tau$-visons that describe $Z_2$ vortices of $\tau$, and
$[\sigma\tau]$-visons that describe a composite of $\sigma$ and $\tau$
$Z_2$ vortices. The three are not independent, any one of them can be
thought of as a composite object of the other two. However, we should
treat all of them on equal footing since they represent distinct
topological objects.  The appearance of the long range order in the
Ashkin-Teller model corresponds to the condensation of visons in the
original gauge model and describes transition to the confining
phase. From these arguments it follows that there are five distinct
phases of the pure gauge model in (\ref{SG}). One fully confining
phase, three partially confining phases, and one fully deconfining
phase, that correspond to the fully ordered, three partially ordered,
and one fully disordered phases of the Ashkin-Teller model.
\begin{trivlist}
\item 1)  Fully confining phase. $\sigma$- and $\tau$-visons are condensed simul

taneously. 
This also  implies condensation of $[\sigma\tau]$-visons.
\item 2)  Partially confining phase. $\tau$-visons are condensed and
$\sigma$- and $[\sigma\tau]$ visons are gapped.
\item 3)  Partially confining phase. $\sigma$ visons are condensed and
$\tau$ and $[\sigma\tau]$-visons are gapped.
\item  4) Partially confining phase.  $[\sigma\tau]$ visons are condensed and
$\sigma$ and $\tau$ visons are gapped.
\item  5) Deconfining phase. All  visons are gapped
\end{trivlist}

Condensation of visons has dramatic effects on the motion of spinons,
holons, and neutral fermions in the model (\ref{S})-(\ref{SG}).  We find
drastically different excitation spectra depending on what vortices
are condensed.  The reason for this is a geometrical phase factor of $\pi$
that particles with $Z_2$-charges acquire when they circle around an
appropriate $Z_2$ vortex. For example, spinons and neutral fermions
get a geometrical phase factor of $\pi$ when they are transported around
a $\sigma$-vison, and holons and neutral fermions get a minus sign
when they circle around a $\tau$-vison. This means that when
visons are present in the ground state, the coherent motion of the
corresponding particles is highly frustrated and they
may not be considered as elementary excitations.
Only the particles that are neutral with respect
to the appropriate $Z_{2}$ symmetry may propagate freely in a phase
with condensed  visons. And the particles that carry such  $Z_{2}$ charges
will have to bind into neutral pairs. This is the essence of the
confinement argument discussed in \cite{Fradkin79,Senthil99}.

When we apply the geometrical phase - confinement argument  
to the spin singlet insulating states 
we find the same phases as discussed in Section \ref{EN}.
\begin{itemize}
\item In a phase of type 1) all kinds of $Z_2$-vortices are
condensed. Therefore, particles that carry any $Z_2$ charges will be
bound. This is a fully confining phase where only fully neutral
composites are allowed. Holons, spinons and neutral fermions are confined.
Phase {\it CSf}.
\item In a phase of type 2) we have a condensate of $\tau$-visons. 
As a result particles that carry $Z_{2\tau}$ charges
are confined, but particles that carry $Z_{2\sigma}$ charges
are liberated. Spinons are free, and
holons are bound to the neutral fermions. Phase {\it CfSb}.
\item In a phase of type 3), that has a condensate of $\sigma$-visons,
we have a confinement of particles with $Z_{2\sigma}$ charges 
and deconfinement of particles with $Z_{2\tau}$ charges. Holons are free, and
spinons are bound to the neutral fermions. Phase {\it CbSf}.
\item In a phase of type 4) we do not have individual $\sigma$- and $\tau$-
visons in the ground states, but only their composites,
$[\sigma\tau]$-visons. The geometrical phase argument becomes somewhat 
subtle when we consider 
$[\sigma\tau]$-visons. Particles that carry either one of $Z_{2\tau}$
or $Z_{2\sigma}$ charges will get a $\pi$ phase shift when they circle around
such a vortex. However, particles that carry both charges acquire no phase.
So, in  a D-type phase particles that carry one of the $Z_{2\tau}$
or $Z_{2\sigma}$ charges are confined, but particles that carry both charges
are deconfined. Holons and spinons
are bound, and neutral fermions are free. Phase {\it CSbNf}.
\item 
Finally in a phase of type 5) we have no condensed visons, which means that
all the particles are liberated. Holons, spinons, and neutral fermions
are deconfined. Phase {\it CbSbNf}.
\end{itemize}

\subsection{Broken Symmetry Phases}
 
In this section we show using $Z_2 \times Z_2$ theory that even states
with the long range order in the model (\ref{S})-(\ref{SG}), i.e.  $p$ wave
superconductors, spin singlet superconductors, nematic insulators,
and nematic superconductors
may differ in their topological ordering and carry the remnants of the
spin-charge separation that appears so dramatically in the insulating
phase.

We begin by reviewing the case of a $p$ wave superconductor. 
\begin{itemize}

\item 
The simplest $p$ wave superconductor that may be deduced from the
model (\ref{S}) - (\ref{SG}) is when holons and spinons condense
simultaneously, so
the system acquires finite expectation values of $b$ and
$z$.  The geometrical phase argument when applied to
this system tells us that an isolated $hc/2e$ vortex or a meron are no
longer well defined excitations, since they acquire a phase shift of
$\pi$ when circling around a holon or a spinon respectively.
However, if we bind an $hc/2e$ vortex with a $\tau$-vison we find that
this composite can propagate freely. The geometrical phases acquired by
the two upon encircling a holon add up to $0$ or $ 2
\pi$. Equivalently, a meron, when bound to a $\sigma$-vison, becomes a
well defined excitation in the presence of spinon condensate. 

\item Another possible phase of a $p$-wave superconductor is when
we condense holon pairs and spinons, i.e. $b^2$ and $z$. 
In this phase merons are still bound to 
$\sigma$-visons, however $hc/2e$ vortices and $\tau$-visons are
now  deconfined. The original holons are reduced to Ising
variables, which we can call $b$-isons, following \cite{Senthil99}.
They carry the leftover of the charge symmetry, that was broken from
$U(1)$ to $Z_2$, and are well defined excitations in this phase.

\item  Analogously to the previous case we can consider a situation
with condensed $b$ and $z^2$. This phase will have bound 
$hc/2e$ vortices and $\tau$-visons and liberated merons and $\sigma$-visons.
Spinons become Ising variables, $z$-isons, that carry the residual 
$Z_2$ spin quantum numbers

\item A different  type of a $p$-wave superconductor
occurs when holon pairs and holon-spinon composites 
condense simultaneously, i.e.
$b^2$ and $bz$ acquire expectation values (this also fixes the 
expectation value for $z^2$).  In such a phase spinons and holons
are reduced to a single Ising variable, since knowing $b$ automatically
gives $z$. This $bz$-ison carries the residual spin-charge quantum
number of the system. A stable topological object 
in this phase may be constructed by taking any two of the set
($hc/2e$ vortex, meron, $\tau$-vison, $\sigma$-vison). 

\item Finally, we may have a phase with condensed holon and spinon 
pairs, $b^2$ and $z^2$. 
This gives us separate $b$-isons, $z$-isons, $hc/2e$ vortices,
merons, $\tau$-visons, and $\sigma$-visons.

\end{itemize}
The last four phases are the triplet analogues of the exotic SC$^*$ phase
discussed in  \cite{Senthil99} in the case of singlet superconductors.

We now consider the case of a spin-singlet superconductor. 
\begin{itemize}
\item
The simplest kind of a spin singlet superconductor occurs when we condense
simultaneously holons $b$ and $\sigma$-visons. The former ensures
confinement of $hc/2e$ vortices and $\tau$-visons, whereas the latter gives
rise to binding of neutral fermions to  spinons. 
\item Another possibility is to have a condensate of holon pairs $b^2$ and
$\sigma$-visons. This liberates $hc/2e$ vortices and $\tau$-visons,
produces $b$-isons that carry charge $Z_2$ number, and leaves neutral
fermions bound to spinons.
\item Another option is to have a condensate of bosons $b$ with gapped
$\sigma$-visons. This means bound $hc/2e$ vortices and $\tau$-visons,
and liberated neutral fermions and spinons.
\item The most intriguing phase in this series is obtained when we
condense holon pairs $b^2$ and holon-$\sigma$-vison composites.
Excitations in this phase will be any pair from the set 
($hc/2e$ vortex, $\tau$-vison, $\psi$, $z$) and $b$-isons.
\item Finally we can have a condensate of $b^2$ and gapped $\sigma$ visons.
This gives unconfined $hc/2e$ vortices, $\tau$-visons, neutral
fermions, spinons, and $b$-isons.
\end{itemize} 
Of the five phases above, four of the last ones may be considered as SC$^*$ phas

es.

The construction given above for $p$-wave superconducting states and
spin singlet superconducting states may be generalized to the case of
spin-nematic insulators and nematic superconductors. In those cases,
just as in the two discussed above, we find five possible states. One
of these is a traditional version, whereas the other four are of the
unconventional *-variety that may be thought of as containing traces of
quantum number separation.

The reader may be worried that
we do not find $hc/4e$ -vortices-$\pi$-disclinations in our discussion
of various phases of $p$-wave superconductors.  As in the previous
sections we assumed that these excitations have been gapped
out (see discussion after equation (\ref{refeq})).

\section{Distinguishing Different Fractionalized Phases}
\label{ddfp}

In previous sections, we have seen how
various fractionalized phases can arise in the context
of Kondo lattice models and systems with a tendency towards
$p$-wave superconductivity or superconductivity coexisting
with magnetism. These phases can be described in the
language of vortex and skyrmion condensation or
in terms of a ${Z_2}\times{Z_2}$ gauge theory.
However, one might wonder if these results are
an artifact of these formalisms. In particular,
one can ask how these phases can be distinguished -- both
as a matter of principle and as a practical experimental
issue -- from each other and from unfractionalized phases.
As Wen \cite{Wen90} and, more recently,
Senthil and Fisher \cite{topth} have emphasized recently,
their `topological order'
-- i.e. the sensitivity of the ground state to
changes of the topology of the system --
provides one means of distinguishing fractionalized
phases.

This characterization of fractionalized phases is
crucial because other heuristic definitions of
fractionalized phases can fail. To see why this is
so, consider the intuitively-appealing
statement that a fractionalized phase is 
distinguished from a conventional phase by asking for
the lowest energy excitation with,
for instance, spin-$1/2$. In the conventional case,
this would be an electron which also
carries an electric charge $e$. In the fractionalized
phases of the kind discussed 
above, one might expect that the corresponding excitation
is a spinon which is charge
neutral. However, this test for fractionalization fails
if there is an attractive interaction between the
holons and spinons which binds them into an
electron at low energies. This could,
in principle, happen without going through a phase transition.
(Unlike in an unfractionalized phase, holons and spinons would
still exist as unbound excitations, but at higher energies.)
Then, the lowest energy excitation with spin-$1/2$
is an electron (as opposed to a spinon) though the 
system is adiabatically connected to a fractionalized phase (see Ref. 
\cite{Balents99,Balents98,Balents00}
for a discussion of this effect). Furthermore, other tests such
as the vanishing of the quasiparticle residue at some point of the Brillouin
zone also fail in this situation. 
Hence, we turn to the characterization in terms of the topological
properties of the system.

Topologically ordered systems are partially
characterized by their ground state degeneracy
on the annulus, the torus, or higher-genus manifolds,
over and above any degeneracy which may be due
to broken symmetry. Consider the CbSf phase.
It has a two-fold degenerate ground state on the annulus.
The two ground states correspond to periodic
and antiperiodic boundary conditions for holons
and spinons as they encircle the center of the annulus.
In either case, electrons themselves have periodic boundary conditions,
as they must. In an unfractionalized phase, spinons and holons are
confined within an electron so the two states are identical;
the excitations which could distinguish them are not
part of the spectrum. By the same reasoning,
the CfSb and CSbNf phases also
have two degenerate ground states on the annulus.
By extension, all of these states have ground state
degeneracy $4^g$ on a genus $g$ surface.
On the other hand, CbSbNf has four degenerate ground states.
We can independently choose periodic or antiperiodic boundary
conditions for the charge and spin bosons. The boundary
conditions for the neutral fermions are then determined by the
requirement that electrons must have periodic boundary conditions.
On a genus $g$ surface, it has degeneracy $16^g$ .

These degeneracies can be interpreted in terms
of the vison spectra of the fractionalized states.
The two ground states of CbSf on an annulus
correspond to the presence or absence of a
$\tau$ vison (i.e. a $v$) in the center of
the annulus; the two ground states
of CfSb correspond to the presence or absence
of a $\sigma$ vison (a $v'$);
the two ground states of CSbNf correspond
to the presence or absence
of a $\sigma\tau$ vison in the center of
the annulus. The four ground states of
CbSbNf correspond to the presence or absence
of $\sigma$ and $\tau$ visons in the center of the annulus.
The interpretation of these ground states in terms
of visons forms the basis for an experimental probe
of topological order proposed by Senthil and Fisher
\cite{toexp}. We will return to this issue later
but let us, in the meantime, continue to pursue the
question of the distinction in principle
between different fractionalized phases.

Different states at the same level of fractionalization have
the same ground state degeneracy; CbSf, CfSb, and CSbNf all
have two degenerate ground states on the annulus.
In order to distinguish them, we must consider their
quantum number spectra. CSbNf does not have spin-charge separation,
i.e. it is not possible to isolate a charge-0, spin-$1/2$
excitation at finite energy cost. Furthermore, it is possible to
isolate a neutral fermionic excitation. Both of these
stand in contrast to CbSf and CfSb which exhibit spin-charge
separation but do not support neutral fermionic
excitations. Hence, we conclude that CSbNf is distinct
from the other two states despite having the same ground
state degeneracy.

One might be tempted to conclude that CbSf and CfSb are distinct
because the lowest-energy charged excitation is a boson in
one phase and a fermion in another phase. However, if
a holon in CbSf forms a bound state with a $\tau$ vison,
the resulting bound state will be fermionic; similarly,
if a spinon in CbSf forms a bound state with a $\tau$ vison,
the resulting bound state will be bosonic. Hence, as
a result of the seemingly innocuous formation of bound states,
the CbSf and CfSb states appear to morph into each other.
Thus one is instead tempted to conclude that the CbSf and CfSb
phases can be adiabatically connected to each other.

This contention is supported by considering the
singlet superconducting state which results if
holons condense in CbSf or if holon-$\sigma$ vison composites
condense in CfSb (see Figure \ref{sfig}). 
It is easy to see that the superconducting states in either case 
are conventional and are smoothly connected to a BCS state. 
The superconducting
state can be disordered by vortex condensation. This will
yield a fractionalized state (with a two-fold degenerate ground state
on an annulus) if vortex pairs condense but individual
vortices are uncondensed. 
Since the result could be either
CbSf or CfSb, this  appears to support the possibility that 
there is no phase boundary between these phases in the
part of the phase diagram near the singlet
superconducting phase.

However, there is a logically-possible alternative,
namely that an operator which is irrelevant
in the superconducting phase and at the critical point
becomes relevant at the fixed points
characterizing the fractionalized phases.
In that case, the actual nature of the resulting
fractionalized phase depends on short distance
physics -- the value of the coupling which is
formally irrelevant in the superconductor --
and is not uniquely dictated by knowing that
there is proliferation of $hc/e$ and with
$hc/2e$ vortices remaining gapped. 

Despite this caveat, a scenario in which CbSf and CfSb are
smoothly connected to each other 
in the vicinity of their transition to the superconducting
state is appealing and plausible.
This does not necessarily mean that CbSf and CfSb
are not distinct phases. Their relationship
could be similar to that between a liquid
and a gas, which are separated by a first-order phase transition
line which terminates at a critical point, beyond which
a liquid and a gas can be adiabatically connected without crossing
a phase transition line. In appendix C, we show
that precisely such a scenario
does occur in simpler (though somewhat different)
${Z_2}\times{Z_2}$ gauge theory
models. Thus, we tentatively suggest that the
first-order phase transition between the
$CbSf$ and $CfSb$ phases terminates at a critical
point. Beyond this critical point, there is no
distinction between these phases, and it is in this
region of the phase diagram that there is
a phase transition to the superconducting phase.

\section{Flux-Trapping Experiments}
\label{cfp}

Let us now consider the practical issue of
how we can identify whether a given system in
an unknown phase is fractionalized or not
and, if it is fractionalized, then what
its fractionalization pattern is.

To proceed, note first that the $CbSf$ phase contains
in it the seed of superconductivity. 
As argued in Ref. \cite{Senthil99}, condensing the
charged boson provides a natural
non-pairing route to superconductivity (of a conventional kind). 
Similarly, the $CfSb$ phase contains in 
it the seed of magnetism - simply condensing the spinon
leads to a conventional state with some
kind of magnetic long range order.
However, it is possible to imagine a transition
between the CfSb phase and a superconductor
which occurs when a composite formed by
a holon and a $\sigma$ vison condenses.
Similarly, it is possible to imagine
a transition between the CbSf phase and
a magnetic phase which occurs when a composite
formed by a holon and a $\sigma$ vison condenses.

The feature of most interest to the following discussion
is simply that a direct phase transition should be
possible between the $CbSf$ and $CfSb$ phases
and a conventional superconductor. Upon going
through such a phase transition, the visons of
these phases acquire $hc/2e$ units of electromagnetic
flux to become the  $hc/2e$ vortices of the superconductor.
This may be exploited to devise a sensitive test for the 
topological order in the $CbSf$ phase, as argued in
Ref. \cite{toexp,topth}.

The test proceeds as follows.
Consider an annular sample of a material which is
in a conventional superconducting phase and let us suppose
that we can tune the sample parameters adiabatically
so that the sample makes transitions between the
superconducting phase and the CbSf and CfSb phases.
Suppose that $hc/2e$ units of electromagnetic
flux are trapped in the annulus when the system is
in its superconducting phase. There must also be
a vison trapped in the annulus so that the holon condensate
can have periodic boundary conditions (without which
it would cost infinite energy): the antiperiodicity caused
by the flux $hc/2e$ is cancelled by the antiperiodicity
due to the vison. If the system is taken
into the CbSf phase, the flux escapes since there
is no holon condensate trapping it,
but a vison will remain since it will cost energy
(the vison gap) to unwind the antiperiodic
boundary conditions of the (neutral) spinons.
If the system is returned to
the superconducting state, then it must generate
flux $\pm hc/2e$ so that the holon condensate
can again have periodic boundary conditions.
The same analysis holds if we take the system
into the CfSb phase except that we have to replace `holon'
in the above description by `holon-vison composite'.
On the other hand, if the system undergoes a transition to
an unfractionalized phase, then the
vison can escape since there are no deconfined
spinons or holons whose boundary conditions would
be affected by its escape.

Of course, this experiment would simply be confirming the
result which we arrived at in the previous section:
that the CbSf and CfSb phases can be adiabatically continued
into each other, particularly in the neighborhood
of a conventional singlet superconducting phase.

Let's now consider a more complicated flux-trapping
experiment in which, as an intermediate step,
we take the system through the higher-level fractionalized phase,
CbSbNf (see Figure \ref{sfig}). 
This phase has two distinct vison excitations. One of these 
visons can be envisioned as a descendent of
the $\tau$ vison of the $CbSf$ phase; we
will refer to this as $v$. The other can be
envisioned as a descendent of
the $\sigma$ vison of the $CfSb$ phase
or as a by-product of the further fractionalization
of the fermionic spinon of $CbSf$; we
will refer to this as 
$v'$.  A direct transition from $CbSbNf$ to the $CbSf$
phase occurs when the visons $v'$ condense
while that from $CbSbNf$ to $CfSb$ occurs when the visons $v$ condense. 
The presence of two distinct visons in the $CbSbNf$ phase
distinguishes it from the $CbSf$ and 
$CfSb$ phases - indeed it will have a ground state
degeneracy of $16$ on a torus.  

Now consider a conventional BCS superconductor. 
This is obtained from $CbSf$ by condensing the holon.
The flux-trapping experiment performed
by moving between the superconductor and $CbSf$ gives a
positive result. Now consider a
modification of the experiment so that
we start in the superconducting phase, move first to 
$CbSf$, then to $CbSbNf$, then back into $CbSf$ before finally 
going back into the superconductor. This again gives a positive result
since $v$ is trapped in the annulus and it can never
escape. Upon making the transition between the CbSf
and CbSbNf phases, a $v'$ will be generated with
probability $1/2$ since the ground state of CbSf
with one $v$ will make a transition to either of the
corresponding ground states of CbSbNf with equal probability.
However, this $v'$ will escape upon the transition
from $CbSbNf$ back to $CbSf$.
Now consider a further modification in which we go all the
way from the superconductor 
to the $CfSb$ phase through the $CbSf$ and $CbSbNf$
phases and then return by the same 
route to the superconductor.
The result of this experiment will be negative
{\em half of the time}. This is because
in going from $CbSbNf$ to $CfSb$, the vison $v$
condenses. Thus $v$ which was 
trapped in the hole until the phase $CbSbNf$ was
reached can escape on moving into the $CfSb$ 
phase. In going from $CfSb$ back to CbSbNf,
a $v$ is generated with probability $1/2$ --
the two ground states are obtained with equal
proability. This $v$, if it is generated,
will lead to the generation of flux $hc/2e$
in the superconducting state.

\begin{figure*}[h]
\epsfysize=7.0cm 
\epsfbox{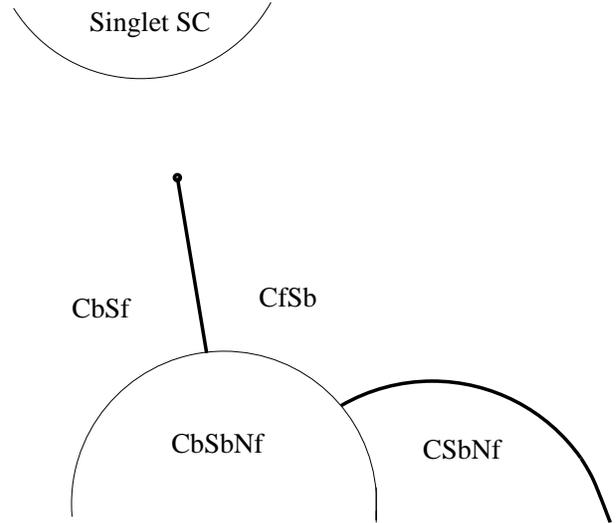}
\caption{A schematic phase diagram indicating how
the CbSf, CfSb, CSbNf, CbSbNf, and conventional singlet
superconducting phases might fit together. The thick lines
are first-order phase transition and the thin lines
are second-order phase transitions.}
\label{fig:1st-order}
\end{figure*}

Hence, there appears to be a difference between
the $CbSf$ and $CfSb$ which can be detected
in this experiment. It appears that these phases
cannot be continuously connected -- since the
probability of a negative result for the
flux-trapping experiment of the previous paragraph
must jump from $0$ to $1/2$ -- at least in the
vicinity of the CbSbNf phase. This can be understood
in the following terms. In the CbSbNf phase, there are two
distinct types of visons, $v$ and $v'$. If one
or the other condensed, a transition occurs to
CfSb or CbSf. The remaining vison in CbSf `remembers' that
it is a $v$ vison. Meanwhile the vison in CfSb
`remembers' that it is a $v'$ vison. However, if
we take the system far from CbSbNf so that
a bound state can form between a $v$ and a holon
and also between a $v$ and a spinon, then $v$
now looks like a $v'$ and the distinction between the two phases
is blurred. Combining this reasoning with that
of the previous section, we propose the phase
diagram of figure \ref{fig:1st-order}.

\section{Discussion}

When electrons interact strongly, a number
of interesting phenomena are known to occur,
including 
unconventional superconductivity
and magnetism. As we have seen in this paper,
many of the physical settings which give
rise to these phenomena also have the potential
to exhibit electron fractionalization.
Different theoretical approaches,
adapted to these specific systems, suggest
seemingly different fractionalized phases.
It is natural to ask if these phases are truly
different and, if so, what their organizing principle is.

In this paper, we have pursued the idea\cite{Wen1,topth} that
a crisp and coherent way of
understanding quantum number fractionalization is provided by the
concept of {\it topological order} introduced in the context of
fractional quantum Hall liquids \cite{Wen3}, anyon superfluids
\cite{anyon}, chiral spin states \cite{chiralspin}, and short range
resonating valence bond spin states \cite{RVB}.  We presented two
approaches for understanding such topological order.  The first one
relies on the recently developed $Z_2$ gauge theory of spin-charge
separation, originally suggested for the high $T_c$ cuprates, and
generalizes it to a $Z_2\times Z_2$ theory to include possible
fractionalization of spin and charge quantum numbers.
Some of the interesting fractionalized phases are: 
CbSf (bosonic holons and fermionic spinons),
CfSb (fermionic holons and bosonic spinons), CSbNf (bound bosonic
holons and spinons and neutral fermions), and CbSbNf (bosonic holons and
spinons and neutral fermions). Any one of these phases
can be further characterized by possible broken symmetries
with conventional order parameters.
Each of the fractionalized phases 
corresponds to a different deconfining
phase of the pure $Z_2\times Z_2$ gauge theory
and will have appropriate topological $Z_2$ vortices, visons, as
finite energy excitations. 

An alternative picture of
fractionalization which is also
presented in this paper uses the language of quantum
disordered superconductors and magnets.
When topological ordering -- defined by the
suppression of certain defects -- occurs,
the Goldstone modes associated with
various broken symmetries can screen the
corresponding quantum numbers of the
fermionic quasiparticles. In this way,
these quasiparticles can be bleached of
some or all of their quantum numbers. This
may be implemented mathematically with
$U(1)$ particle-vortex duality
in both the charge and spin sectors. We arrive
at essentially the same picture as that of the
$Z_2\times Z_2$ gauge theory.
In those insulating phases in which $hc/2e$ vortices
are condensed, charge is bound to the fermionic
quasiparticles. When $hc/2e$ vortices are gapped and 
$hc/e$ vortices are condensed,
charge carrying "holons" can propagate separately from the
electrically neutral fermionic quasiparticles. In the spin sector,
we can consider either meron or skyrmion (which carry twice the topological
charge of merons) condensation, with gapped
merons in the latter case. In the former case, spin
is confined to the fermionic
quasiparticles, and in the latter case "spinons" will
exist as independent objects, deconfined from the fermionic
quasiparticles. We have also discussed the possibility of quantum disordered
phases in which the condensed topological objects are the $hc/2e$ vortex  -
meron composites, but not $hc/2e$ vortices or merons separately. Such
phases have "spinons" and "holons" bound together but deconfined from
the neutral fermionic quasiparticles.

An important issue discussed in this paper is whether one can
distinguish the phases obtained by 
quantum disordering the spin and charge sectors
of the system, for example the phases
CfSb and CbSf of the quantum disordered $p$-wave superconductor.
The simplest choice seems to be the identification
of the spin excitation as a fermionic or bosonic particle.
This, however, is not a reliable tool.
In the $Z_2\times Z_2$ gauge theory formulation,
both spinons and holons carry
$Z_2$ charges, so a bound state of a
$Z_2$ vortices with either one of them
(this can also be thought of as attaching Wilson loops to the particles)
will change its statistics from fermionic to bosonic or vice
versa \cite{Read89,Ng99a,Ng99b,Wilczek}. 
In the deconfining phase such vortices are gapped.
However, if a bound state between
a $Z_2$-charge carrying particle and a $Z_2$
vortex forms, this bound state
may have a lower energy than the
original particle. This means that in both {\it CfSb} and 
{\it CbSf} phases the
lowest energy spin- or charge-carrying excitations can exist as 
either bosons or fermions.
The subtleties discussed above lead us to consider
flux-trapping experiments
of the type discussed in Section \ref{cfp}.
Combining all of these considerations, we outlined one scenario in which 
{\it CbSf} and {\it CfSb} phases can be separated by
a first order transition which terminates at a critical
point. On the other hand, one can go from {\it CbSf} to
{\it CfSb} through {\it CbSbNf} phase by two continuous
transitions. Thus, if this scenario is correct, the relation between {\it CbSf} 
and 
{\it CfSb} is somewhat similar to that between
liquid and gas phases. We however defer offering any definitive conclusion.

Spin charge separation in one dimensional
systems is fundamentally different from
its two dimensional counterpart, since it does
not involve topological order. 
Another nontrivial realization
of electron number fractionalization which is analogous
to that presented here
can occur in multi-component
quantum Hall systems and was discussed
in references \cite{QH1,QH2}.

Another avenue for further research is
the investigation of quantum-disordered
states of triplet superconductors with
more complicated spin structures appearing
in some of the superfluid phases of $^3$He. 
We expect that these will
share some features of non-collinear
spin-density-waves \cite{Chubukov94}.
Further exotic phases are likely
to occur upon quantum-disordering
states with multiple order parameters.
We have considered one of the simplest
cases of this -- antiferromagnetism
and superconductivity -- but there are
more complicated possibilities,
involving incommensurate charge
and/or spin order.

In addition to the phases CbSf and CfSb that have appeared in the literature bef
ore, 
we proposed the possibility of two additional
quantum number separated phases in these systems: phase CSbNf in
which the excitations are a spin-$1/2$, charge $e$ boson, a neutral spinless fer
mion and a vison
and phase CbSbNf with a charge $e$ spinless boson, a neutral spin $1/2$ boson, a
 neutral spinless fermion, and 
{\em two} distinct visons. 
simultaneously condensed $hc/e$ vortices and skyrmions.

The possibility of a higher SO(5) symmetry which unifies d-wave 
superconductivity and antiferromagnetism
has been suggested for the high
$T_c$ cuprates and organic superconductors in \cite{Zhang97}. 
In the $Sr_2 Ru O_4$ materials, a similar symmetry
has been proposed in
\cite{Murakami99} which combines $p$-wave
superconductivity and ferromagnetism. 
An effective model for the coupling of
quasiparticles to a fluctuating SO(5) order parameter has been
derived in \cite{Demler99}. In this model holons and spinons are not
segregated into independent quasiparticles from the very beginning but are
naturally combined into composite quasiparticles which transform as
spinors of SO(5). Such spinors are spin doublets and carry charge $e$
\cite{Rabello98,Scalapino98}.
There are also neutral fermions which
carry no quantum numbers.  One can see a striking resemblance between
these excitations and the excitations in the phase CbSbNf. 
This suggests the interesting possibility
that the restoration of the SO(5) symmetry in models with strong 
quantum fluctuations manifests itself not in the existence
of a bicritical point on the phase diagram, 
but in the appearance of a specific form of quantum number
separation of the electrons. A detailed discussion of
quantum disordering phenomena in models with SO(5) symmetry
requires a detailed analysis of the non-Abelian Berry's phases
involved in the description of SO(5) spinors and 
will be presented in subsequent publications.

The states which we have discussed,
as well as the more complicated
ones alluded to above, have potential application
to a variety of materials,
including not only the cuprates \cite{Bednorz},
but also ${Sr_2}{Ru}{O_4}$ \cite{Maeno}; heavy fermion
superconductors, such as $CeIn_3$ \cite{Stewart};
and organic superconductors, such as
$\kappa-(ET)_{2} Cu[N(CN)_{2} ]Cl$ \cite{organic}.
All of these compounds
have magnetic (in come cases incommensurate)
phases in proximity to $p$-wave or
$d$-wave superconducting states. It is possible that
pressure, chemical substitution, magnetic field, etc.
might drive a transition into one of the phases
described here in which the magnetism and
the superconductivity are disordered by
quantum fluctuations.

Ideas presented in this paper should also apply to Bose-Einstein
condensates of spinor bosons, such as 
alkali atoms $^{23}Na$ and $^{87}Ru$ which have a hyperfine spin
$F=1$. For example, when restricted dimensionality or 
quantum fluctuations
destroy the spin ordering we expect to find condensation 
of pairs of atoms into a global spin singlet state, and
when quantum fluctuations in the charge sector 
destroy the $U(1)$ phase ordering we can find states characterized by a
spin nematic order.
Some of these phenomena have been discussed in 
\cite{Zhou}.

To summarize, we have studied the possibility of 
fractionalization in systems with ordering tendencies
in the charge and spin
sectors, including Kondo lattices, $p$-wave superconductors
and systems with simultaneous 
$d$-wave superconducting and antiferromagnetic fluctuations. In the case of
$p$-wave superconductors we find that the rich internal structure of their
order parameter allows for the existence of the following quantum disordered
phases: a charge $4e$ singlet superconductor, a spin singlet insulator,
and a spin nematic insulator. 
For both the $p$ wave superconductors and the $d$-wave
superconductor/antiferromagnet systems, we find that
the quantum disordered phases may have separated
quantum numbers, depending on the
topological order, which can be characterized by specifying
the  nature of the finite energy $Z_2$ visons.

\acknowledgements
ED and CN thank Aspen Center for Physics for its hospitality
during the Winter 2000 Conference ``50 Years of
Condensed Matter Physics'',
where some parts of this work were initiated. 
HYK and YBK thank ITP, University of California at Santa Barbara
where some parts of this work were performed.
CN, HYK, and YBK thank Aspen Center for Physics for its
hospitality during the summer workshop in 2000.
We also thank Aspen Center for Physics for its
hospitality during the winter workshop in 2001.
Useful discussions with M.P.A. Fisher, E. Fradkin, S. Kivelson, and
M. Sigrist are gratefully acknowledged. 
This work was supported by the Harvard Society of Fellows (ED);
NSF under grant numbers DMR-9983544 (CN) and 
DMR-9983783 (YBK); the Alfred P. Sloan Foundation (CN and YBK);
the Department of Energy, supported (in part) by funds provided
by the University of California for the conduct of discretionary
research by Los Alamos National Laboratory (HYK). The work of TS 
at the ITP, Santa Barbara, was supported by the NSF under Grants DMR-97-04005,
DMR95-28578
and PHY99-07949.

\appendix
\section{Kondo lattice model}
In this appendix, we provide some of the details of the $Z_2$ gauge 
theory reformulation of the 
Kondo lattice model discussed in Section \ref{sss}. 
Consider the Hamiltonian in 
Eqn. \ref{kondo}.
As in the discussion of the pure exchange Hamiltonian,
we first replace the spin 
operator $S^-_r$
by the boson operator $b_{sr} \sim e^{i\varphi_r}$.
The exchange Hamiltonian takes the 
form Eqn. \ref{hexbos} and the Kondo coupling takes the form Eqn. \ref{kb}. 
The electron hopping term is unaffected. We now change variables to spinon and 
holon operators
as in Eqns. \ref{splitsp}, \ref{splitcua} and \ref{splitcda}.
The terms $H_t, H_k$ 
and $H_{ex}$ are now given by 
Eqns. \ref{htcs}, \ref{ks}, and \ref{hexsp} respectively.
In the presence of the 
Kondo coupling between the local moments
and the conduction electrons, the total ($z$ component of the)
spin at each site is 
\be
n_r + \frac{1}{2}c^{\dagger}_r  \sigma^z c_r
\ee
We therefore define the total spinon number
\be
N^{tot}_r = 2n_r + c^{\dagger}_r \sigma^z c_r
\ee  
Note that $N^{tot}_r$ is conjugate to the phase $\phi_r$ of the spinon field. 
We will work with the operators
$(z_r, N^{tot}_r, \eta_{\ua r}, \eta_{\da, r})$ 
instead of the original electron and local spin $\vec S_r$ operators. This 
change of variables however introduces some redundancy -
the Hilbert space of states
on which the holon and spinon fields operate is larger
than the physical set of states. 
This may be seen by noting that with the definition
above, the operator $N^{tot}_r$
must satisfy
\be
N^{tot}_r - c^{\dagger}_r \sigma^z c_r = even
\ee
>From the definition
of the holons, it follows that 
$c^{\dagger}_r \sigma^z c_r = \eta^{\dagger}_r \sigma^z \eta_r$.
Furthermore
$ \eta^{\dagger}_r \sigma^z \eta_r$ has the same
parity as $ \eta^{\dagger}_r  \eta_r$. 
Thus we have the constraint
\be
N^{tot}_r - \eta^{\dagger}_r  \eta_r = even
\ee
The Hamiltonian needs to be supplemented with this 
constraint to correctly represent the original model (before the change of 
variables). 

It is useful to rewrite the exchange and Kondo parts
of the Hamiltonian as follows:
 
\bea
H_K + H_{ex} & = & J_K \sum_r \left (\eta^{\dagger}_{r\ua}\eta_{r\da} 
+ h.c. \right) \nonumber \\
& & -J \sum_{<rr'>} \left(z^{2\dagger}_{r}z^2_{r} + h.c \right)
+ \frac{U}{4}\sum_r(N_r - 1)^2
\nonumber \\
& & -U\sum_r N_r (\eta^{\dagger}_r \sigma^z \eta_r)
+ \frac{U}{4} \sum_r (\eta^{\dagger}_r \sigma^z \eta_r)^2
\eea

The last term is an interaction between the holons.
Clearly this term cannot affect 
issues of confinement of the holons with the spinons.
We will therefore drop it
for the present discussion. The last but one term represents an 
interaction between the spinon density and the holons.
We again expect that such an interaction is also unimportant
for issues of the 
stability of fractionalized phases.
We will therefore drop 
this too. 

We may now derive a functional integral representation
of the system, proceeding as in
Ref. \cite{Senthil99}. 
The resulting action is 
\be
S = S_{\tau} + S_r + S_B
\ee
Here $S_{\tau}$ represents terms involving coupling along the (imaginary)
time direction.  
  This and the Berry phase $S_B$ are
exactly the same as in Ref. \cite{Senthil99}. The spatial part of the 
action is 
\bea
S_r & = & S_I + S_K + S_{II} \\
S_I & = & -\epsilon \sum_{<rr'>} t_{rr'} \left(z^{\dagger}_{r} z_{r'} + 
\left(\eta^{\dagger}_{r'\ua}\eta_{r\ua} + 
\eta^{\dagger}_{r\da}\eta_{r'\da} \right) + h.c. \right) \nonumber \\
S_K & = & + \epsilon J_K \sum_r \left(\eta^{\dagger}_{r\ua}\eta_{r\da} 
+ c.c.\right) \nonumber \\
S_{II} & = & -\epsilon J \sum_{<rr'>}
\left(z^{2\dagger}_{r}z^2_{r} + h.c. \right) 
\eea
We now combine the terms $S_I$ and $S_{II}$ and rewrite them as
\bea
& & -\epsilon J \sum_{<rr'>} \left(z^{\dagger}_{r} z_{r'} + \frac{t_{rr'}}{2J} 
\left(\eta^{\dagger}_{r'\ua}\eta_{r\ua} +
\eta^{\dagger}_{r\da}\eta_{r'\da} \right)\right)^2 \nonumber \\
& & + h.c. + {\cal O}(\eta^4)
\eea
The last term is a four-holon interaction which
we will ignore on the grounds that it
cannot affect issues of fractionalization. It is
convenient to further rewrite the
expression above as follows:
\bea
S_{I} + S_{II} & =  & \frac{\epsilon J}{2}
\sum_{<rr'>}\left[\left(z^{\dagger}_{r}z_{r'} + 
\frac{t_{rr'}}{2J}\eta^{\dagger}_r \eta_{r'} +
h.c. \right)^2 \right. \nonumber \\
& & + \left. \left(z^{\dagger}_{r}z_{r'} - 
\frac{t_{rr'}}{2J}\eta^{\dagger}_r \eta_{r'} - h.c. \right)^2  \right]
\eea
We may now decouple each of these two terms
with a {\em real} Hubbard-Stratanovich field
to write
\bea
& & e^{-(S_I + S_{II})}  =  \int [{\cal D}\chi
{\cal D}\rho] e^{-\left(S_{\chi} + S_{\rho} \right)} \\
& & S_{\chi}  =  \frac{\epsilon J}{2}\sum_{<rr'>} \chi_{rr'}^2 -2\chi_{rr'} 
\left(z^{\dagger}_{r} z_{r'} + 
\frac{t_{rr'}}{2J}\eta^{\dagger}_r \eta_{r'} + h.c. \right) \\
& & S_{\rho}  = \frac{\epsilon J}{2}\sum_{<rr'>} \rho_{rr'}^2 -2\rho_{rr'} 
\left(z^{\dagger}_{r}z_{r'} - 
\frac{t_{rr'}}{2J}\eta^{\dagger}_r \eta_{r'} - h.c. \right)
\eea
Note that $\chi_{rr'} = \chi_{r'r}$ while
$\rho_{rr'} = -\rho_{r'r}$. We now consider
evaluating the $\chi, \rho$ integrals in a saddle point approximation.  
Looking for uniform saddle points, we write 
\be
<\chi_{rr'}> = \chi_0;~~~<\rho_{rr'}> = \rho_0
\ee
Note that a non-zero value of $\rho_0$
requires specifying directions for all the links
of the lattice. 
The saddle point equations are
\bea
\chi_0 & = & <z^{\dagger}_{r}z_{r'} + 
\frac{t_{rr'}}{2J}\eta^{\dagger}_r \eta_{r'} + h.c. > \\
\rho_0 & = & <z^{\dagger}_{r}z_{r'} - 
\frac{t_{rr'}}{2J}\eta^{\dagger}_r \eta_{r'} - h.c. >
\eea
Note that $\rho_0$ must be pure imaginary as it is the expectation 
value of an antihermitian operator. 
With non-zero $\rho_0$, the saddle point action therefore becomes 
{\em complex} - this breaks time-reversal symmetry (and possibly various 
lattice symmetries due to the need to specify directions to the links). 
We restrict ourselves to time-reversal 
invariant saddle point solutions, and therefore set $\rho_0 = 0$. 
The resulting saddle point action then preserves all the global 
symmetries of the original model. 
However, it does break the local $Z_2$ symmetry introduced by the 
change of variables to the holons and the spinons. This can be 
remedied by keeping a particular set of fluctuations about the saddle 
point, namely those associated with a change in the sign of 
the fields $\chi_{rr'}$:
\be
\chi_{rr'} = \chi_0 \sigma_{rr'}
\ee
with $\sigma_{rr'} = \pm 1$. The $\sigma_{rr'}$ may be
identified as the spatial 
components of a $Z_2$ gauge field. We thus finally
arrive at the action in Eqn. 
\ref{z2kon}.

\section{Quantum Number Separation in Systems with 
$d$-wave Superconducting and Antiferromagnetic Fluctuations}
\label{dSCAF}

\begin{figure*}[h]
\centerline{\epsfxsize=6.0cm 
\epsfbox{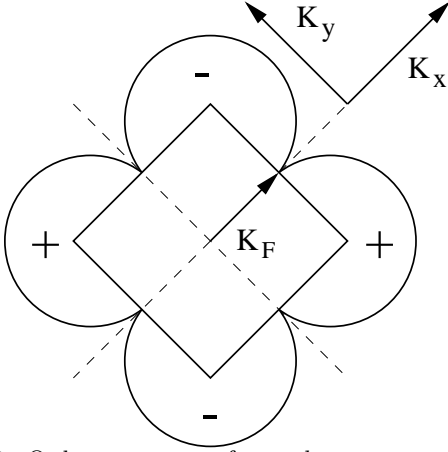}}
\caption{Order parameter for a $d$-wave superconductor. Gapless excitations
exist at $\vec{k}_F= (\pm k_F,0),\,(0,\pm k_F)$.}
\label{dwave}
\end{figure*}

Nodal fermions in a
$d$-wave superconductor are described by \cite{Balents98}
\begin{eqnarray}
{\tilde{S}_f}  &=&  \int {d^2}x\,d\tau\,\chi^\dagger
[ \partial_\tau - {A_\tau}{\tau^z}
 - {v_F} {\tau^z}
  i\partial_x + {v_F} {A_x}\cr
& & \,\,\,\,\,-  A^{\sigma}_{\tau} \vec{n} \vec{\sigma} + 
v_f A^{\sigma}_x \vec{n} \vec{\sigma} \tau_z
-  v_{\scriptscriptstyle\Delta} \tau^s
\left\{ e^{is\varphi}\,\right\}
  (i \partial_y) ] \chi
\label{tildeSf}
\end{eqnarray}
where the electron operators $\chi_{a\alpha}$ are defined as
\begin{eqnarray}
\chi_{a\alpha}(\vec{k}) 
= \left[ \begin{array}{c}
\chi_{11} \\ \chi_{21} \\ \chi_{12} \\ \chi_{22} 
\end{array} \right] 
= \left[ \begin{array}{c}
c_{\vec{ k}_F+\vec{k}\uparrow}^{\vphantom\dagger} \\ 
c_{-\vec{ k}_F-\vec{k}\downarrow}^{\dagger} \\ 
c_{\vec{ k}_F+\vec{k}\downarrow}^{\vphantom\dagger} \\ 
-c_{-\vec{ k}_F-\vec{k}\uparrow}^{\dagger} 
\end{array}\right].
\end{eqnarray}
and the coordinate system was rotated in such a way 
that the $x$ axis goes along the nodal direction
that we are considering ( see Figure \ref{dwave}).

 Antiferromagnetic fluctuations
are introduced via 
\begin{eqnarray}
{S_{af}} &=& \int d\tau\,\int d^2 k d^2 q\,
\vec{n}_{\vec{q}}\, c^{\dagger}_{-\vec{ k}_F+
\vec{k}\alpha} \vec{\sigma}_{\alpha \beta}
c_{\vec{ k}_F+\vec{k}+\vec{q}\beta} ~+h.c.
\nonumber\\
&=&
\int {d^2}x\,d\tau\, \vec{n}(x) \chi_{a\alpha}(x)
 \epsilon_{ab} \epsilon_{\alpha}^{~\gamma} 
\vec{\sigma}_{\gamma\beta} \chi_{b\beta}(x)
\end{eqnarray}
where $\vec{n} = (\cos\theta, \sin\theta)$.

Spin and charge may again be decoupled by rotating the fermions
as in (\ref{psiintro})
\begin{eqnarray}
\label{eqn:bleach}
{\chi_{a\alpha}} = {e^{i\varphi {\tau^z}/2}}\,
{e^{i\theta {\sigma^z}/2}}\,{\psi_{a\beta}}
\label{psiintro1}
\end{eqnarray}
with the result
\begin{eqnarray}
\tilde{ S } &=&  \int {d^2}x\,d\tau\,\psi^\dagger
[ \partial_\tau 
 - {v_F} {\tau^z}
  i\partial_x 
-  v_{\scriptscriptstyle\Delta} \tau^x
  (i \partial_y) ] \psi\cr
& & + \int {d^2}x\,d\tau\,  \psi_{a\alpha}(x) \,
 \epsilon_{ab}\, \epsilon_{\alpha}^{~\gamma} \sigma^x_{\gamma\beta}
\psi_{b\beta}(x)
\nonumber\\
&+&\,\, \frac{1}{2}\int {d^2}x\,d\tau\,\biggl(
\psi^\dagger [{\tau^z} {\partial_\tau}\varphi
- 2{A^c_\tau}{\tau^z}
- {v_F} {\partial_x}\varphi + 2{v_F} {A^c_x}] \psi\cr
& & {\hskip -0.5 cm}\,+\, \psi^\dagger 
[{\sigma^z} {\partial_\tau}\theta
- 2{A^\sigma_\tau}{\tau^z}
- {v_F} {\tau^z}{\sigma^z}
{\partial_x}\theta + 2{v_F} {\tau^z}{\sigma^z}{A^\sigma_x}] \psi
\biggr)
\end{eqnarray}
This describes the same coupling of the quasiparticle
currents  to the fluctuations
of charge and spin as in (\ref{eqn:Gold-qp-int2a})
\begin{eqnarray}
\label{eqn:Gold-qp-afdsc-int2a}
{\tilde{S}_f}  &=& {\tilde{S}^0_f} +
\int {d^2}x\,d\tau\,\biggl(\,
J_0^c ( \partial_{\tau} \varphi - A_{\tau} )
+ J_x^c ( \partial_x \varphi - A_{x} )\cr
& & {\hskip 1 cm}+ J_0^{\sigma} ( z^{\dagger} \partial_{\tau} z
- A^{\sigma}_{\tau} )
+ J_x^{\sigma} ( z^{\dagger} \partial_x z - A^{\sigma}_{x} )
\biggr)
\end{eqnarray}
with 
\begin{eqnarray}
\begin{array}{ll}
J_0^c = \psi^\dagger \tau^z \psi \hspace{1.5cm} &
J_x^c = - v_F \psi^\dagger  \psi  \\
J_0^{\sigma} = \psi^\dagger \sigma^z \psi \hspace{1.5cm} &
J_x^{\sigma} = - v_F \psi^\dagger  \sigma^z \tau^z\psi  \\
\end{array}
\end{eqnarray}
Quantum disordering of the superconducting and antiferromagnetic
orders in (\ref{eqn:Gold-qp-afdsc-int2a}) 
may now be achieved by condensing vortices and merons
with the possibility of five phases
similar to phases 3a-3e in Section \ref{condensation}
\begin{trivlist}
\item A) Spinons and holons confined. No quantum number separation.
\item B) Spinons unbound and holons glued to fermions.
\item C) Holons free and spinons bound to fermions.
\item D) Spinons and holons bound together, decoupled from fermions.
\item E) All excitations decoupled. Free holons, spinons, neutral fermions.
\end{trivlist}

\section{Adiabatic continuation between different phases
of a ${Z_2}\times{Z_2}$ gauge model with matter fields.}

In the pure $Z_2\times Z_2$ gauge theory there are 5 phases that are 
distinct and separated
by phase transitions. A question that we address in this section
is whether this distinction survives in the presence of matter
fields.

\subsection{Toy Models}

\begin{figure}
\epsfxsize=3.5in \centerline{\epsffile{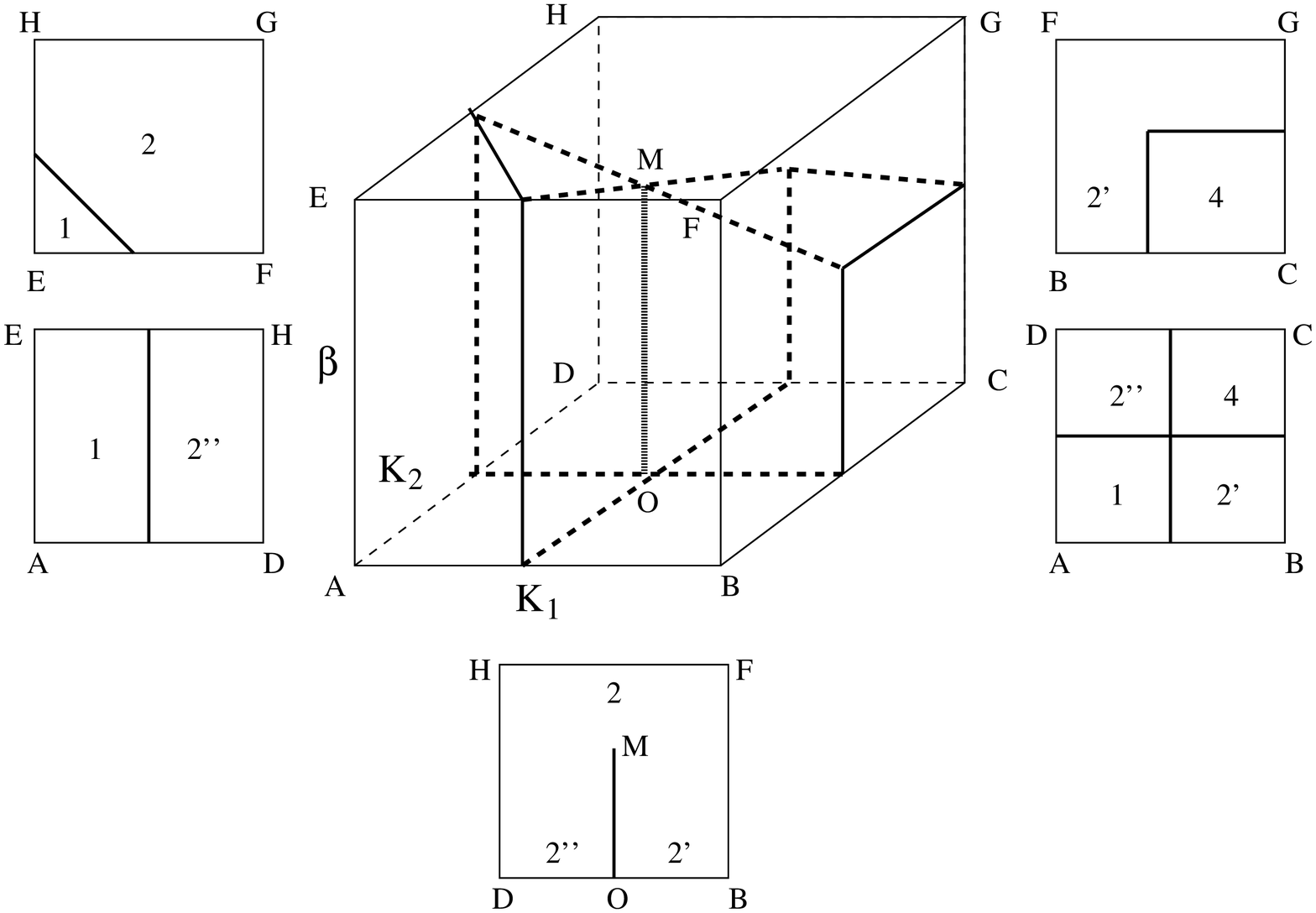}} 
\caption{Phase diagram of (\ref{stv}).
Plane ABCD corresponds to $\beta=0$,
EFGH to $\beta=\infty$,
ADHE to $K_1=0$,
BCFG to $K_1=\infty$, 
ABFE to  $K_2=0$, and
DCGH to $K_2=\infty$.
The pure gauge model ($\beta=0$)
has only two partially confining
phases (2' and 2'') because not all the
possible lattice Maxwell terms
are present.
They are separated by one first order
or two second order phase transitions
for small $\beta$, but 
may be continuously
connected for larger $\beta$.
}
\label{stvfig}
\end{figure}

Let us begin with a simple model
\begin{eqnarray}
S= - K_1 \sum_\Box \prod_\Box \sigma_{ij}
   - K_2 \sum_\Box \prod_\Box \tau_{ij}
   - \beta \sum_{ij} \sigma_{ij} \tau_{ij} v_i v_j
\label{stv}
\end{eqnarray}
where $v_i = \pm 1$ is an Ising matter field.
To construct the full phase digram of this model
we consider several limiting cases. 
When $\beta=0$
we have two independent gauge fields, each
of which has confining and deconfining phases.
When both $K_1$ and $K_2$ are small we have
a confining phase for both $\sigma$ and $\tau$,
that has no extra degeneracy on topologically non-trivial manifolds
and is labeled 1 on Fig. \ref{stvfig}.
When $K_1$ is large and $K_2$ is small
we have a  phase that is confining for $\tau$, but
deconfining for $\sigma$ (phase $2'$ in  Fig. \ref{stvfig}).
There is an analogous phase  for $K_2$ large and $K_1$  small
(phase $2''$ in Fig. \ref{stvfig})  which is
confining for  $\sigma$ and deconfining for $\tau$.
When both $K$s are large, we have a fully deconfining
phase with degeneracy 4 on a cylinder. 
When $K_1=\infty$ (BCGF plane on fig  \ref{stvfig})
there are no frustrated plaquettes for $\sigma$, so we can choose
a gauge where all $\sigma_{ij}=1$. The model is then
the same as in \cite{Fradkin79}
and its phase diagram can be easily constructed.
When $K_1=0$ (ADHE plane on fig  \ref{stvfig})  
we  find that integrating out $\sigma_{ij}$ and 
$v_i$  only adds a constant to the action for $\tau$
and does not affect the confinement-deconfinement
transition which 
takes place for the same value of  $K_2$, regardless
of the value of $\beta$. 
When $\beta=\infty$ (EFGH plane on fig  \ref{stvfig}) 
we must have 
$\prod_\Box \sigma_{ij} \tau_{ij}=1$
on every plaquette, so we can choose a gauge
where $\sigma_{ij} \tau_{ij}=1$ on every link.
The fields $\sigma$ and $\tau$ are identical
and there are only two phases, a confining
(phase 1) and a deconfining (phase 2) with the 
transition determined by
$K_1+K_2$. 
The full phase diagram may now be obtained by connecting the lines on
the faces of the cube on Fig \ref{stvfig}. It is immediately clear
from this picture that the two partially confining phases which appeared
to be distinct for $\beta=0$ (phases $2'$ and $2''$
in the ABCD plane) may be
continuously connected through a path that takes advantage
of the finite $\beta$ region of the phase diagram.
It is important to realize that our argument for
the existence of a path connecting phases $2'$ and $2''$ does not depend
on the details of how the phase boundaries in Fig \ref{stvfig}
are connected. One can always find a path which begins
in phase $2'$, approaches
face ABFE, goes up to EFGH, crosses to EHDA, and finally comes down
to 2'' without crossing the phase boundaries (this path does not have
to actually be on any of the faces and it may be sufficient to be in
their vicinity).  It is interesting to note that the cross-section
DBFH on Fig \ref{stvfig} the phase diagram looks similar to a
liquid-gas phase diagram, where the two phases $2'$ and $2''$ may be
separated by a 1st-order transition or continuously connected
arond the critical point which terminates the 1st-order line.

The real reason why phases $2'$ and $2''$ of the gauge theory (\ref{stv})
may be connected to each other is that both of them are related
to the Higgs phase (phase 2) for the
$v_i$ matter field. This may be explained by noting
that in such Higgs phase visons
of either $\sigma$ or $\tau$ are forbidden, but their composite is
not, so this phase should be related to the phases where these visons 
are condensed separately (but not simultaneously).
 
Note that this argument may no longer apply
if the matter field carries a quantum number, and
a Higgs phase breaks some continuous symmetry.
Let us, for example, explore the model where the matter field
is an $XY$ order parameter
\begin{eqnarray}
S &=& - K_1 \sum_\Box \prod_\Box \sigma_{ij}
   - K_2 \sum_\Box \prod_\Box \tau_{ij}
\nonumber\\
   &-& \beta \sum_{ij} \sigma_{ij} \tau_{ij} 
        cos (\phi_i - \phi_j)
\label{stv1}
\end{eqnarray}
In this case, the Higgs phase has superfluid order, and therefore is fundamental
ly distinct from confined insulating phases. Thus, we can no longer 
easily claim the equivalence of the two phases
in which either $\sigma$ or $\tau$ (though not both) fields
are confining.
  
We can again attempt to construct a phase diagram
following construction on each of the outside
faces of the cube. The ABCD plane is the same
as in Fig \ref{stvfig}. The BCGF plane
($K_1=\infty$) will now have three phases:
a confining and a deconfining phases
without broken XY symmetry (phases $2''$ an 4), 
and a phase
with broken XY symmetry \cite{Lammert93}.
When $K_1=0$ (ADEH plane) we have 4 phases.
This is obvious from the fact that when we integrate out
$\sigma_{ij}$ we find that $\tau_{ij}$ and 
$cos \phi_i$ are decoupled from each other,
and we have separate order-disorder and confinement-deconfinement
transitions. We therefore find two  superfluid
phases $SF_1$ and $SF_2$ that differ
in their degeneracy on the nontrivial manifolds
(on the cylinder it is 1 for $SF_1$ and 2 for  $SF_2$).
The origin of this extra degeneracy for  $SF_2$
is that it has a finite energy 
topological excitation: a bound state of $\sigma$ and $\tau$ visons
that does not interact with the matter field.
This has  interesting implication that we have $hc/2e$ vortices
that are bound to either $\sigma$ or $\tau$ visons, and
the two kinds of vortices are distinct.

\begin{figure}
\epsfxsize=3.5in \centerline{\epsffile{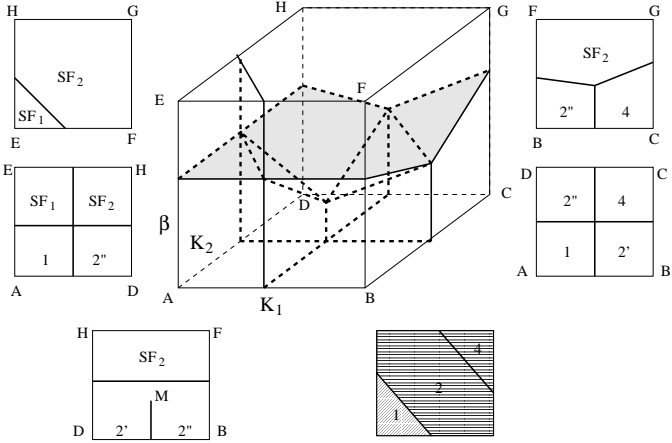}} 
\caption{One scenario for the phase diagram of (\ref{stv1})
with two kinds of superfluid
phases $SF_1$ and $SF_2$. Shaded figure shows
a phase boundary of the superfluid and insulating phases.
Diagonal shading corresponds to the boundary of  $SF_1$
and horizontal shading to a boundary of  $SF_2$.
There is a continuous path to go between
partially confining phases 2' and 2'' without crossing the phase
boundaries.
}
\label{stv1fig}
\end{figure}

We do not at this stage know what the generic phase diagram 
for (\ref{stv1}) in the $K_1 K_2 \beta$ cube is.
One possibility is shown on Figure  \ref{stv1fig}.
As in the Ising case there is 
a way to connect phases 2' and 2'' continuously
by going to finite $\beta$. 

\begin{figure}
\epsfxsize=2in \centerline{\epsffile{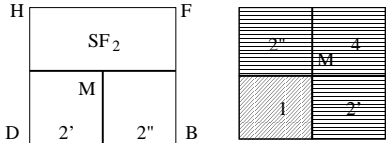}} 
\caption{Another scenario for a  phase diagram of (\ref{stv1})
when the point M is exactly on the superfluid-insulator
phase boundary. As before shaded figure shows
a phase boundary of the superfluid and insulating phases
with diagonal and horizontal striping that correspond
to  $SF_1$ and  $SF_2$ respectively.
In this case there is no continuous path connecting 
phases 2' and 2'' without crossing the phase
boundaries.}
\label{stv2fig}
\end{figure}

There is another qualitatively different phase diagram  (without
introducing  new phases) where the point M is at
the phase boundary with the superfluid phase.
This will remove the possibility of a continuous
path between phases $2'$ and $2''$ (see  Figure  \ref{stv2fig}).

At this point we are unable to make a definite comment on 
of the validity of either of the scenarios shown on Figure  
\ref{stv1fig} or Figure \ref{stv2fig}. We note however that 
this issue is amenable to study by numerical or other 
means. Thus, future work should be able to settle this satisfactorily.

\begin{figure}
\epsfxsize=3.5in \centerline{\epsffile{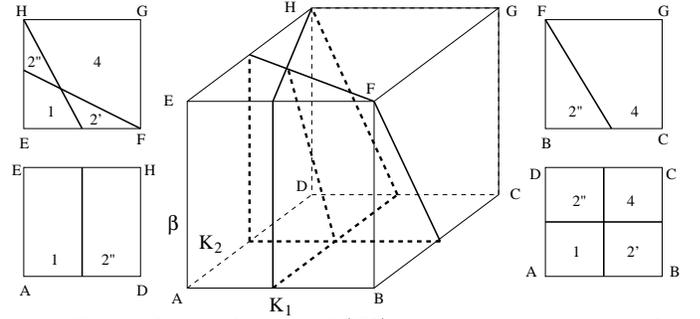}} 
\caption{Phase diagram of (\ref{stv3}). When the matter
field is fermionic distinction between phases 2' and 2''
survives for all $\beta$.}
\label{stv3fig}
\end{figure}

Another important model to consider is one in which
the matter field is fermionic. An appropriate model is 
\begin{eqnarray}
S= - K_1 \sum_\Box \prod_\Box \sigma_{ij}
   - K_2 \sum_\Box \prod_\Box \tau_{ij}
   - \beta \sum_{ij} \sigma_{ij} \tau_{ij} \psi_i \psi_j
\label{stv3}
\end{eqnarray}
where the $\psi$'s are real fermions. Following the same
kind of arguments as before we find the phase diagram shown on
Fig \ref{stv3fig}. There is no Higgs phase for the fermions
which leads to phases 2' and 2'' being distinct even for
finite $\beta$.

\subsection{Full Action}

Let us now consider the  action (\ref{S})
and ask how many truly distinct phases it has.
The phase space of this model is large and an explicit construction
of the full phase diagram is difficult. We note however that the charge sector o
f 
the theory is precisely the same as (\ref{stv1}). Consequently, 
if in $\ref{stv1}$, 
the two phases $2$ and $2'$ are smoothly connected to each other, they will
necessarily be so for the full action as well. If on the other hand, 
in (\ref{stv1}), the two phases are distinct, 
then that is evidence (though not proof) that they 
are distinct 
in the full theory as well. 

Thus, unambiguous determination of the phase diagram of (\ref{stv1}) 
will shed considerable light on the important conceptual issue of whether 
CbSf and CfSb are distinct or not.

While the distinction between CbSf and CfSb, if any, is subtle,
it is very clear that they are both distinct from CSbNf. 
One can not find any vison attachment
that would map the spectrum of $CSbNf$ to either
$CbSf$ or $CfSb$, which proves rigorously that it is
a phase fundamentally distinct from the other two.

To summarize the discussion in this section, we argued that 
studies, numerical or otherwise, of simple models of the form
of Eqn. (\ref{stv1})
should be extremely useful in deciding on the 
issue of whether CbSf and CfSb are distinct quantum phases.
One can, however, prove rigorously that $CbSf$ and $CfSb$
are fundmentally different
from the other partially confining phase of (\ref{S}) $CSbNf$. 
The other phases of (\ref{S}): 
the fully confining phase ($CSf$) and the fully deconfining
phase ($CbSbNf$) will be distinct from any of the partially
confining ones and from each other as may
be seen from their degeneracy on non-trivial manifolds.

\section{Un-fractionalized Phases}

In this paper, we have, for the most part, focussed on
states in which the electron is fractionalized.
However, even the transitions which do not
lead to electron fractionalization are rather
novel. One would ordinarily assume that strong
quantum fluctuations will completely disorder
a $p$-wave superconductor. However, as we pointed out
in section \ref{td}, if $hc/4e$ vortex-$\pi$ disclination
composites are gapped, then the spin symmetry can
be restored without affecting the charge;
alternatively, the superconductivity can be
destroyed without affecting the spin ordering.

Let us consider, first, what happens when flux
$hc/2e$ vortices condense, but no other topological defects
condense. Then the charged degrees of freedom are disordered,
but the spin nematic order parameter should be undisturbed.
Following the arguments of \cite{Nayak00a}, a possible
un-fractionalized spin nematic insulating state
is (in the notation of \cite{Nayak00a})
a triplet $p_x$ density wave:
\begin{equation}
\label{eqn:pDW_def}
\left\langle {\psi^{\alpha\dagger}}(k+Q,t)\,
{\psi_\beta}(k,t) \right\rangle
= {\vec{\Phi}_Q} \cdot
{\vec{\sigma}^\alpha_\beta}\,\,
\sin {k_x}a
\end{equation}
This state is related to the $p$-wave superconducting
state by a `rotation' generated by
\begin{equation}
{O^+} = \int \frac{{d^2}k}{(2\pi)^2}\,\,
{c_\uparrow^\dagger}(k)\,
{c_\downarrow^\dagger}(-k+Q)
\end{equation}
In other words, the triplet $p_x$ density wave
and the $p$-wave superconductor are related
in precisely the same way as a CDW and an $s$-wave
superconductor. In particular, Hamiltonians with
short-ranged interactions can be constructed for which
both states are exactly degenerate; such Hamiltonians
could describe a critical point between these two
states.

In the triplet $p_x$ density wave state,
there is no spin moment (at any wavevector),
since the right-hand-side of (\ref{eqn:pDW_def})
vanishes upon integration over $\vec{k}$.
However, the spin-nematic order parameter,
which may be calculated from (\ref{eqn:pDW_def}),
is non-vanishing:
\begin{equation}
\left\langle {S_i}{S_j}-\frac{1}{3}{\delta_{ij}}{S^2} \right\rangle
= \frac{1}{2}\,{\left|{\vec{\Phi}_Q}\right|^2} 
\,{\rm diag}(2/3,-1/3.-1/3)
\end{equation}
where $S_i$ is the $i$-th component of the total spin of the system.
Hence, this is the natural spin nematic state
which results when a $p$-wave superconductor
is quantum disordered by flux-$hc/2e$ vortex
condensation. The possibility of spin nematic states in the context 
of high Tc cuprates has been proposed 
in \cite{Gorkov}.

When the spin degrees of freedom are
disordered, but the charge remains ordered,
the triplet $p$-wave superconducting order parameter
and the spin-nematic order parameter vanishes;
only the charge-$4e$ order parameter is left.
The condensation of merons causes the fermionic
quasiparticles to be confined to spin.
Once the merons have condensed, the topological quantum number
in the spin sector is no longer well-defined,
so the flux $hc/4e$ vortex-$\pi$ disclination
composites become simple flux $hc/4e$ vortices,
as we would expect for a charge-$4e$ superconductor.
Said differently, the meron condensate
screens the spin topological charge
of the flux $hc/4e$ vortex-$\pi$ disclination
composites, thereby making rendering them
simple flux $hc/4e$ vortices. Remarkably,
by quantum disordering the {\it spin sector} of
a $p$-wave superconductor, we have changed
the {\it charge} of its order parameter, which
may, for example, lead to some unusual
critical behavior of the superconducting transition.


\begin{references}
\bibitem{endnote1} In principle, we should also consider non-singlet
charge-$4e$ superconductors, with 
${\Delta^{4e}}\neq 0$, $Q \neq 0$
but vanishing $\Delta_{\alpha\beta}$.
These occur when the flux $hc/4e$ vortex-$\pi$ disclination
composite becomes unstable and the flux $hc/4e$ vortex
and the $\pi$ disclination exist as separate
stable excitations. These states will be discussed in 
subsequent publications.
\end{references}
\end{document}